\documentclass{article}
\usepackage{spconf,amsmath,graphicx}
\usepackage[sort&compress,numbers]{natbib}

\usepackage{enumitem}
\setlist{nosep, leftmargin=14pt}
\usepackage{xcolor}
\usepackage{multirow, makecell}

\title{Automated Segmentation of Computed Tomography Images with Submanifold Sparse Convolutional Networks}

\name{Sa\'ul Alonso-Monsalve$^1$ ~~ Leigh H. Whitehead$^2$ ~~ Adam Aurisano$^3$ ~~ Lorena Escudero Sanchez\sthanks{Corresponding author email: les44@cam.ac.uk}$^4$}
\address{$^1$ Institute for Particle physics and Astrophysics, ETH Z\"urich, Zurich CH-8093, Switzerland \\ $^2$ Department of Physics, University of Cambridge, Cambridge CB3 0HE, UK \\ $^3$ Department of Physics, University of Cincinnati, Cincinnati, OH 45221, USA \\ $^4$ Department of Radiology, University of Cambridge, Cambridge CB2 0QQ, UK}

\begin{document}
%\ninept
%
\maketitle
\begin{abstract}
Quantitative cancer image analysis relies on the accurate delineation of tumours, a very specialised and time-consuming task. For this reason, methods for automated segmentation of tumours in medical imaging have been extensively developed in recent years, being Computed Tomography one of the most popular imaging modalities explored. However, the large amount of 3D voxels in a typical scan is prohibitive for the entire volume to be analysed at once in conventional hardware. To overcome this issue, the processes of downsampling and/or resampling are generally implemented when using traditional convolutional neural networks in medical imaging. In this paper, we propose a new methodology that introduces a process of sparsification of the input images and submanifold sparse convolutional networks as an alternative to downsampling. As a proof of concept, we applied this new methodology to Computed Tomography images of renal cancer patients, obtaining performances of segmentations of kidneys and tumours competitive with previous methods ($\sim$\,84.6\% Dice similarity coefficient), while achieving a significant improvement in computation time (2-3 min per training epoch).

\end{abstract}

\section{Introduction} 
\label{sec:intro}
Imaging, such as Computed Tomography (CT), is routinely used for the diagnosis and treatment monitoring of cancer patients, usually assessed in a qualitative manner. A more quantitative analysis of such images is desirable and has indeed shown potential in recent years in terms of predicting disease evolution and treatment response \cite{Rundo2022}. However, the accurate delineation of tumours is crucial to perform such analyses, but it is as well a very time-consuming task that requires expert radiologists, and currently represents an impediment from advanced image analysis to be incorporated into the clinical setting. For this reason, automatic tumour segmentation methods are very valuable for the cancer research community \cite{kits19results,Thomas2022}. 

In practice, using whole scans as 3D images is computationally prohibitive without the application of downsampling. For example, a single CT scan typically consists of $O(100)$ images of size $512\times512$ pixels that would represent a total of $\sim$\,26 million voxels. 
The dense representation used in traditional convolutional neural networks (CNNs) is inefficient when many voxels convey no useful information for the classification process. In this scenario, a sparse representation storing the position and value of only meaningful voxels is more suitable, and submanifold sparse convolutional neural networks (SSCNs)~\cite{Graham-2017-submanifold,Graham-2017-3d} were developed specifically to operate on sparse tensors. Such SSCNs have been applied in different contexts in the last couple of years, for detection, classification, and semantic segmentation problems~\cite{Gwak2020, Huang2020, Xie2020}. For example, in the field of high-energy physics, where sparse data are common, SSCNs have been shown to outperform standard CNNs~\cite{Domine-2020-scalable,Adams:2019uqx,NEXT:2020jmz}. In fact, a first example of such networks used in medical imaging has been recently proposed for reconstruction of 3D skulls~\cite{Li2022}. 

Although CT images are not naturally sparse, this imaging modality is a suitable use case for SSCNs. This is possible since pixels in CT are described by the scale of Hounsfield units (H.U.) that represents radiodensity~\cite{HounsfieldNobel}, such that different values are indicative of different tissue types. Therefore, by selecting thresholds in the pixel intensities, we can discard structures not useful for the tumour segmentation task, such as bones or air. In this paper, we propose a methodology based on pixel sparsification and SSCNs for CT segmentations of organs and tumours, studying their feasibility and performance in the context of kidney and renal tumours.
 
\section{Materials and Methods}
The method explained here consists of three steps. Firstly, the input images are prepared and \textit{sparsified} as explained below. An SSCN model is trained and applied to the sparsified images, downsampling them in this first stage so that broad regions of interest (ROI) can be found around kidneys + masses together. These ROIs will then be used to crop the input images in 3D to train and apply a second SSCN in the second stage, that will perform the final segmentation at full resolution (no downsampling) on the cropped sparse images. 

\subsection{Data samples} 
For this study, we used publicly available images of renal cancer patients collected for the 2021 Kidney Tumour Segmentation challenge (KiTS21)\cite{kits21}. This dataset is the largest public collection of images available of patients who underwent partial or radical nephrectomy for suspected renal malignancy, collected between 2010 and 2020, taken in the late arterial phase. It contains a total of 300 CT scans, which formed the training and validation data for the challenge, plus an additional 100 CT scans that are not public, comprising the test dataset of the challenge. The challenge annotations (ground truth segmentations) for kidney and masses (tumours and cysts) were used, in particular the version using a majority vote amongst several readers. Ethical approval was not required, as stated in the license accompanying the open-access dataset. 

\begin{table*}
\centering
\caption{The effect of removing voxels outside of given ranges of Hounsfield units (H.U.) in the sparsification process for CT scans in terms of: the maximum total number of voxels allowed, the fraction of signal (kidney and masses) voxels lost, and the fraction of background voxels removed. Each row represents the optimal range that minimises signal loss for a given maximum number of voxels following a discrete quantitation of (from bottom to top): 1, 2, 3, 4 and 5. The range given in the second row, -30 to 350 H.U., was chosen as a balance between signal and background losses.}
\vspace{5pt}
\begin{tabular}{c | c | c | c | c}
     \textbf{Max voxels (M)} & \textbf{Min (H.U.)} & \textbf{Max (H.U.)} & \textbf{Signal loss (\%)} & \textbf{Bkg loss (\%)}  \\
     \hline
     \rule{0pt}{2ex} 125 & -150 & 350 & 0.10 & 52.86\\
     & & & (0.008, 0.20) & (51.79,53.94)\\
     \hline
     \rule{0pt}{2ex} 64 & -30 & 350 & 2.08 & 76.63 \\
     & & & (1.88,2.28) & (75.92,77.34)\\
     \hline
     \rule{0pt}{2ex} 27 & 50 & 240 & 25.99 & 89.30 \\
     & & & (24.52,27.47) & (88.90,89.70)\\
     \hline
     \rule{0pt}{2ex} 8 & 110 & 230 & 57.66 & 97.10 \\
     & & & (55.55,59.78) & (96.92,97.28) \\
     \hline
     \rule{0pt}{2ex} 1 & 130 & 140 & 94.69 & 99.64 \\
     & & & (94.35,95.04) & (99.61,99.67)\\
     \hline
\end{tabular}
\label{tab:factors}
\end{table*}

\begin{figure}[tbh]
  \includegraphics[width=0.48\textwidth, height=0.7\textwidth]{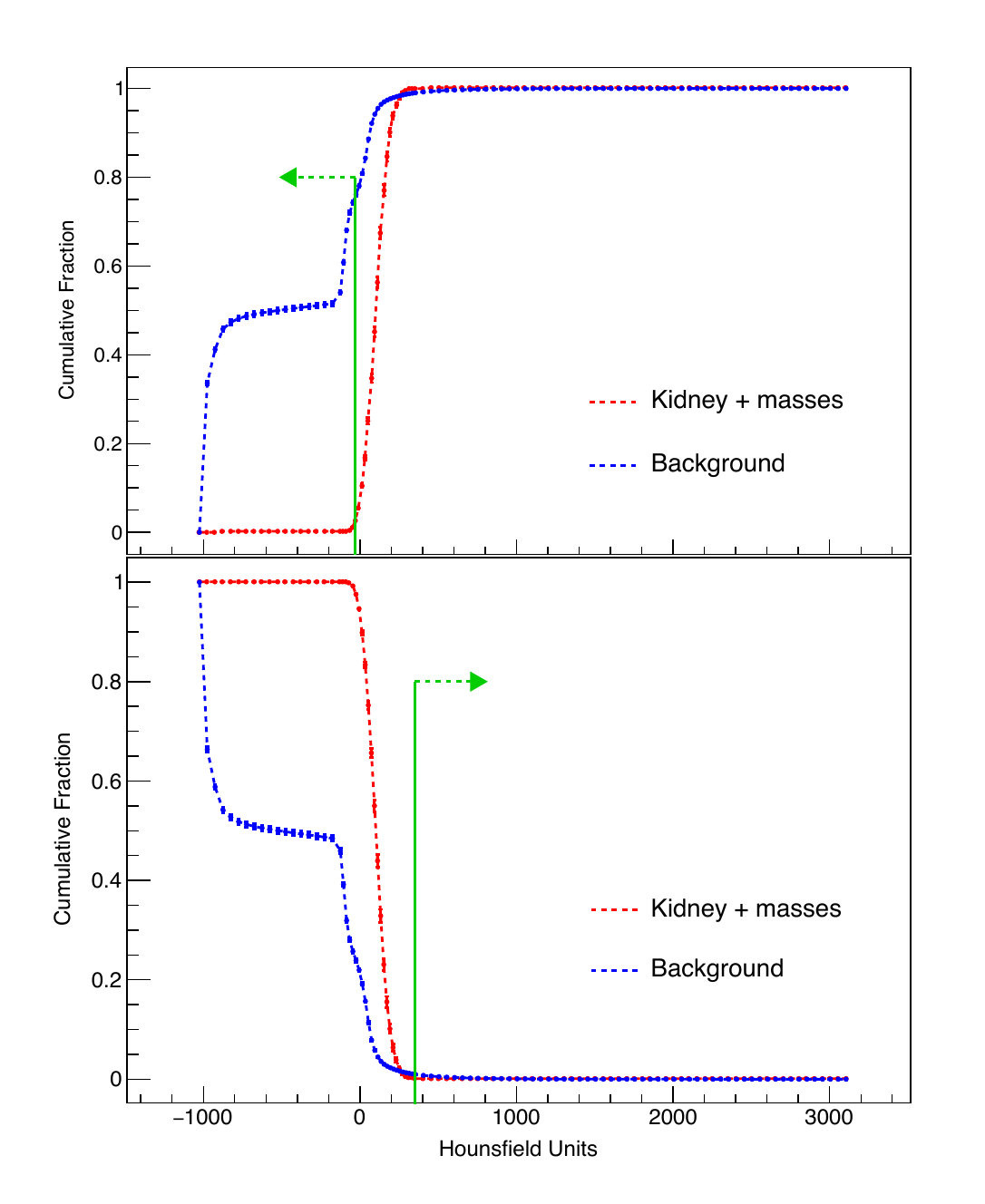}
  \caption{The cumulative fraction of voxels removed by applying a minimum (top) and maximum (bottom) threshold to their intensity in Hounsfield units, shown for the kidney and masses (red) and all other pixels (blue). The green arrows indicate regions rejected in the sparsification process.}
  \label{fig:min_max_pixels}
\end{figure}

\begin{figure*}[th]
  \includegraphics[width=0.33\textwidth,height=0.3\textwidth]{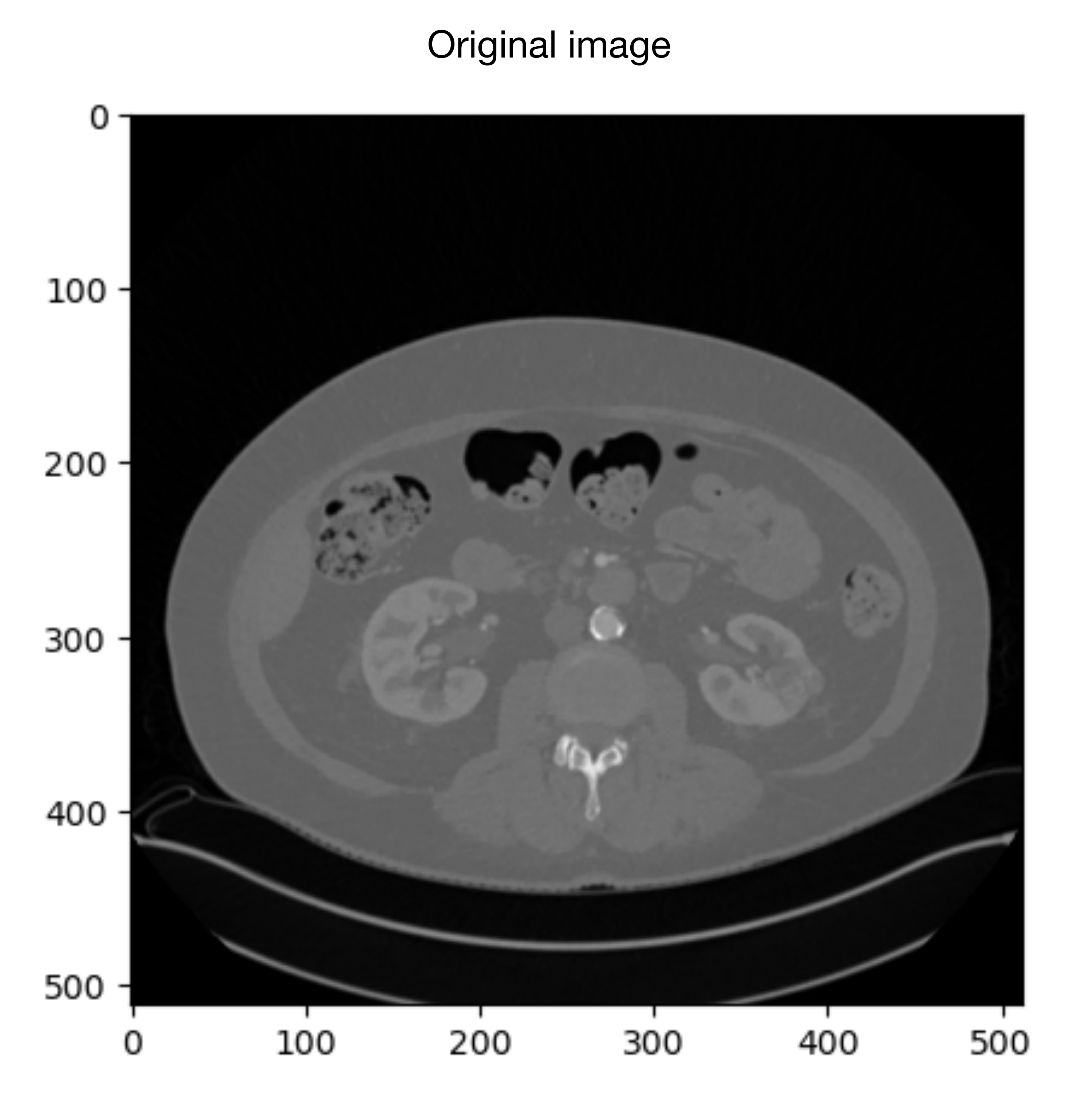}\includegraphics[width=0.33\textwidth,height=0.3\textwidth]{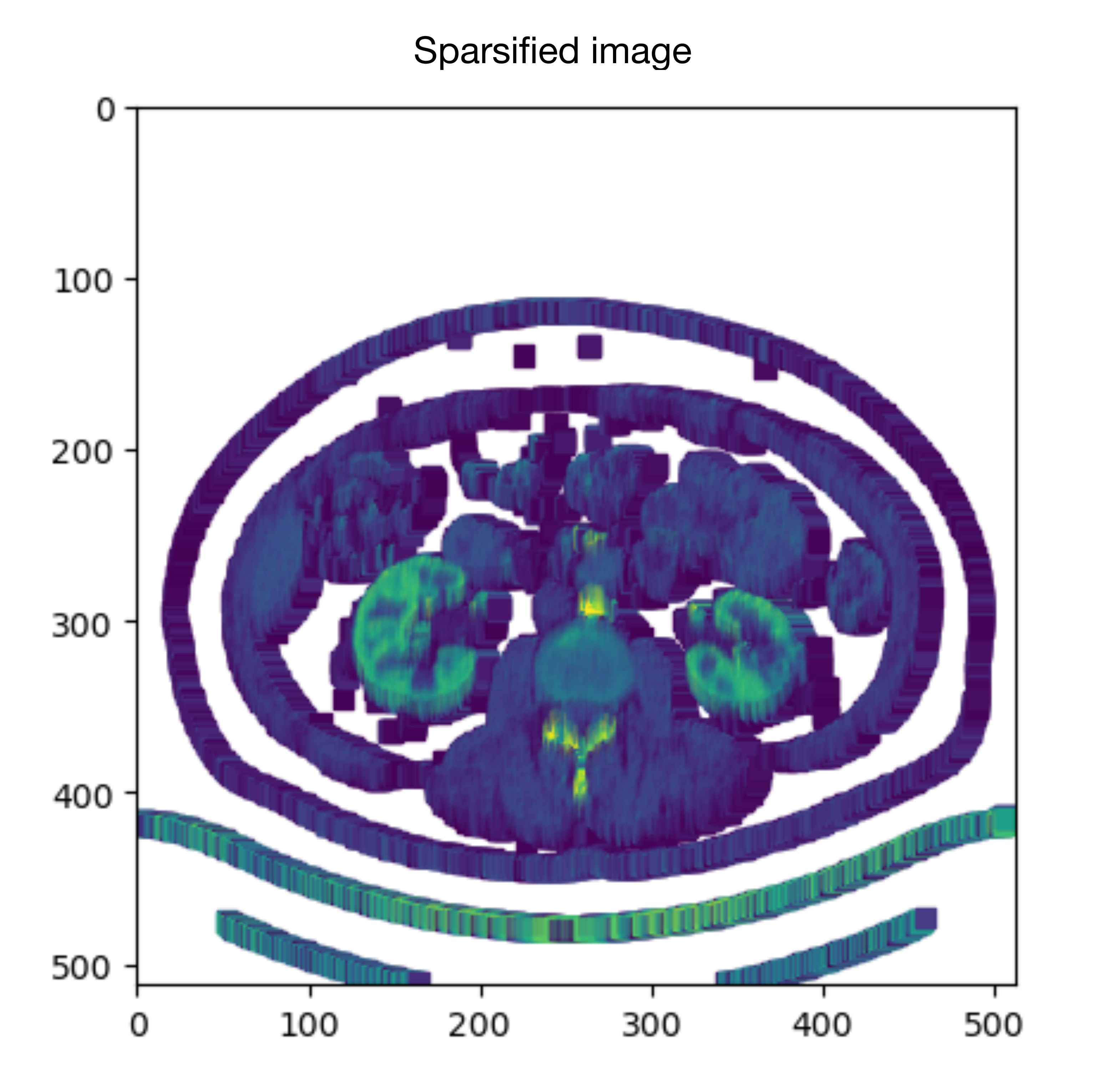}\includegraphics[width=0.33\textwidth,height=0.3\textwidth]{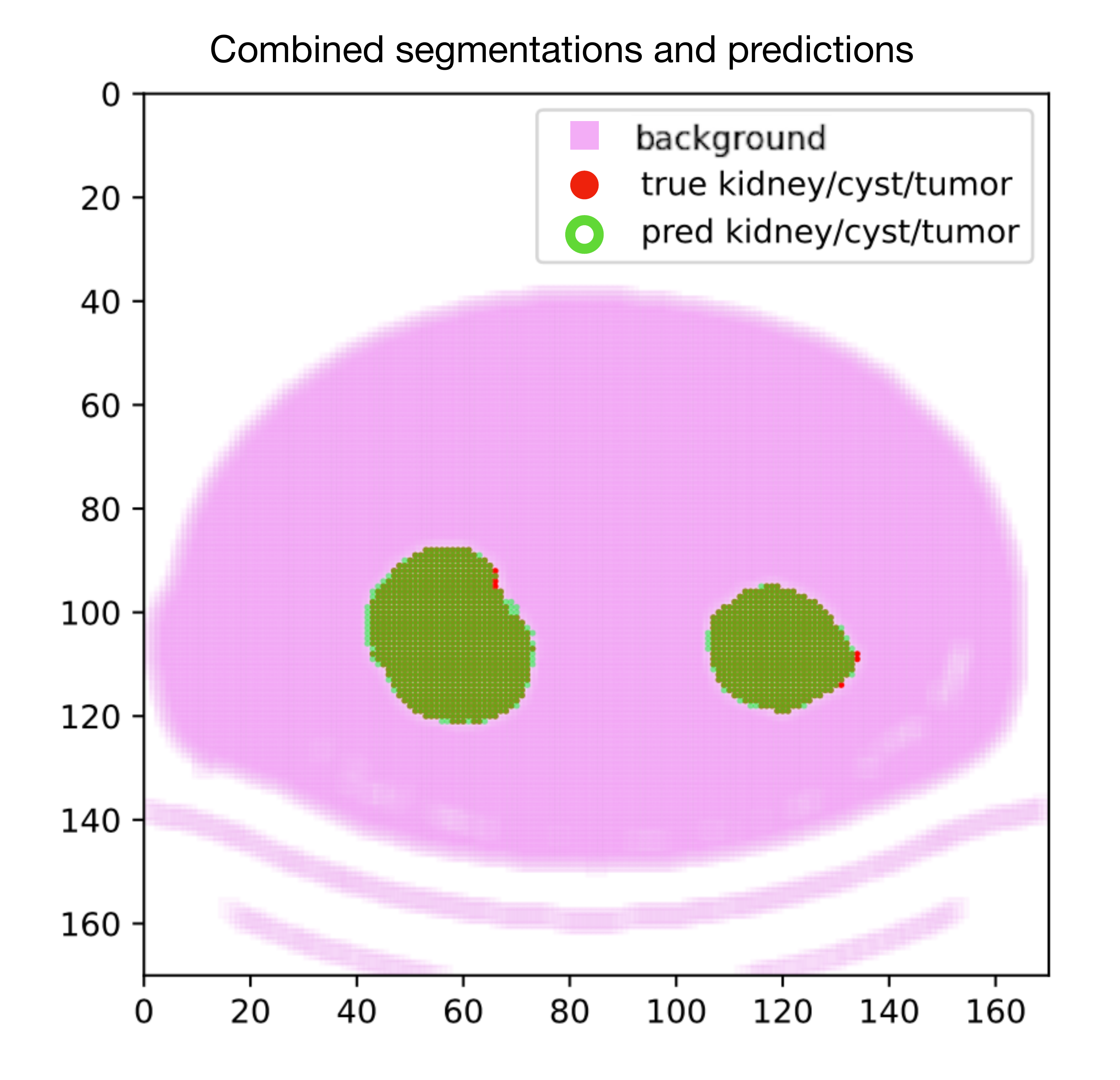}\\
    \includegraphics[width=0.33\textwidth,height=0.3\textwidth]{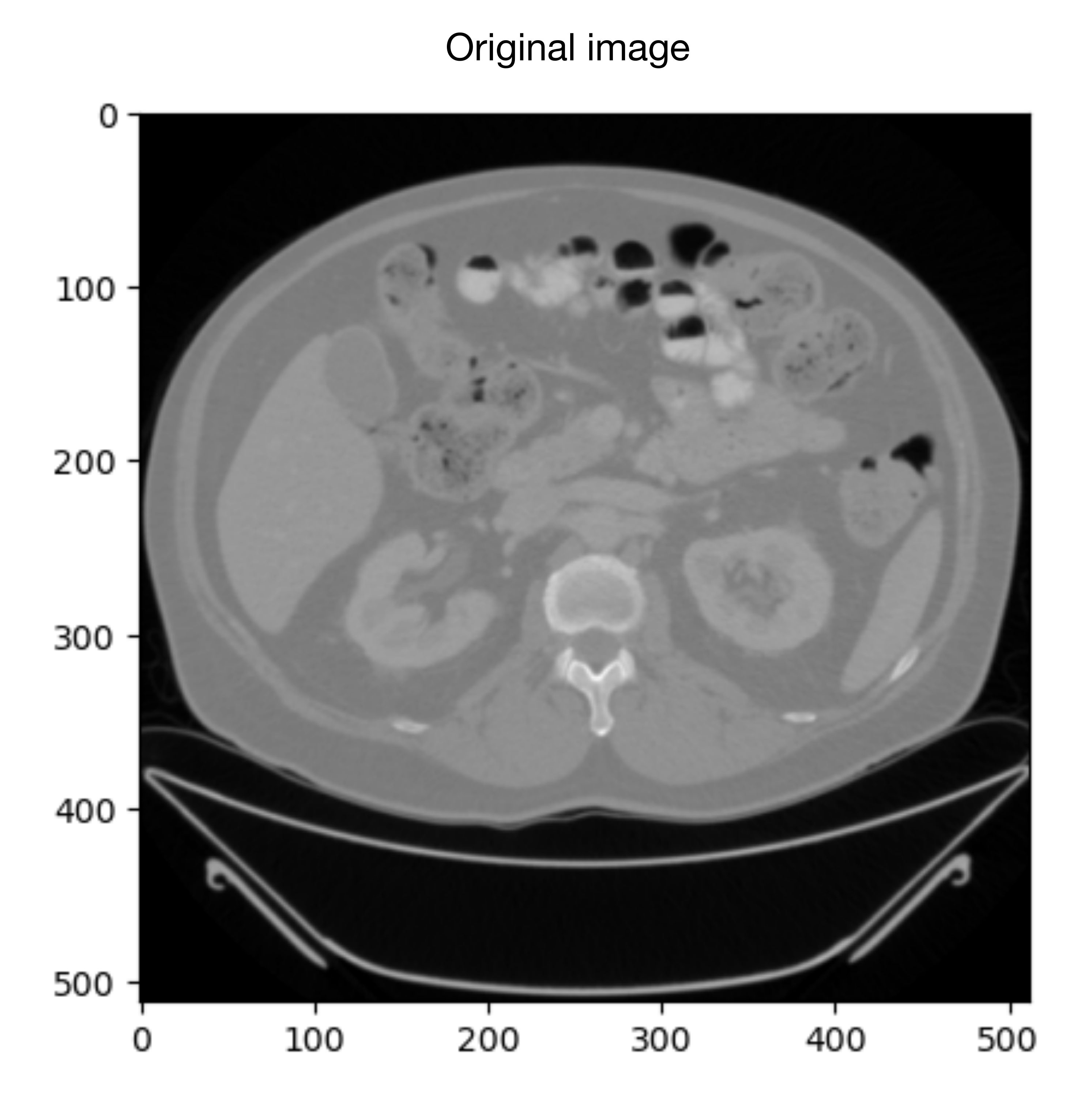}\includegraphics[width=0.33\textwidth,height=0.3\textwidth]{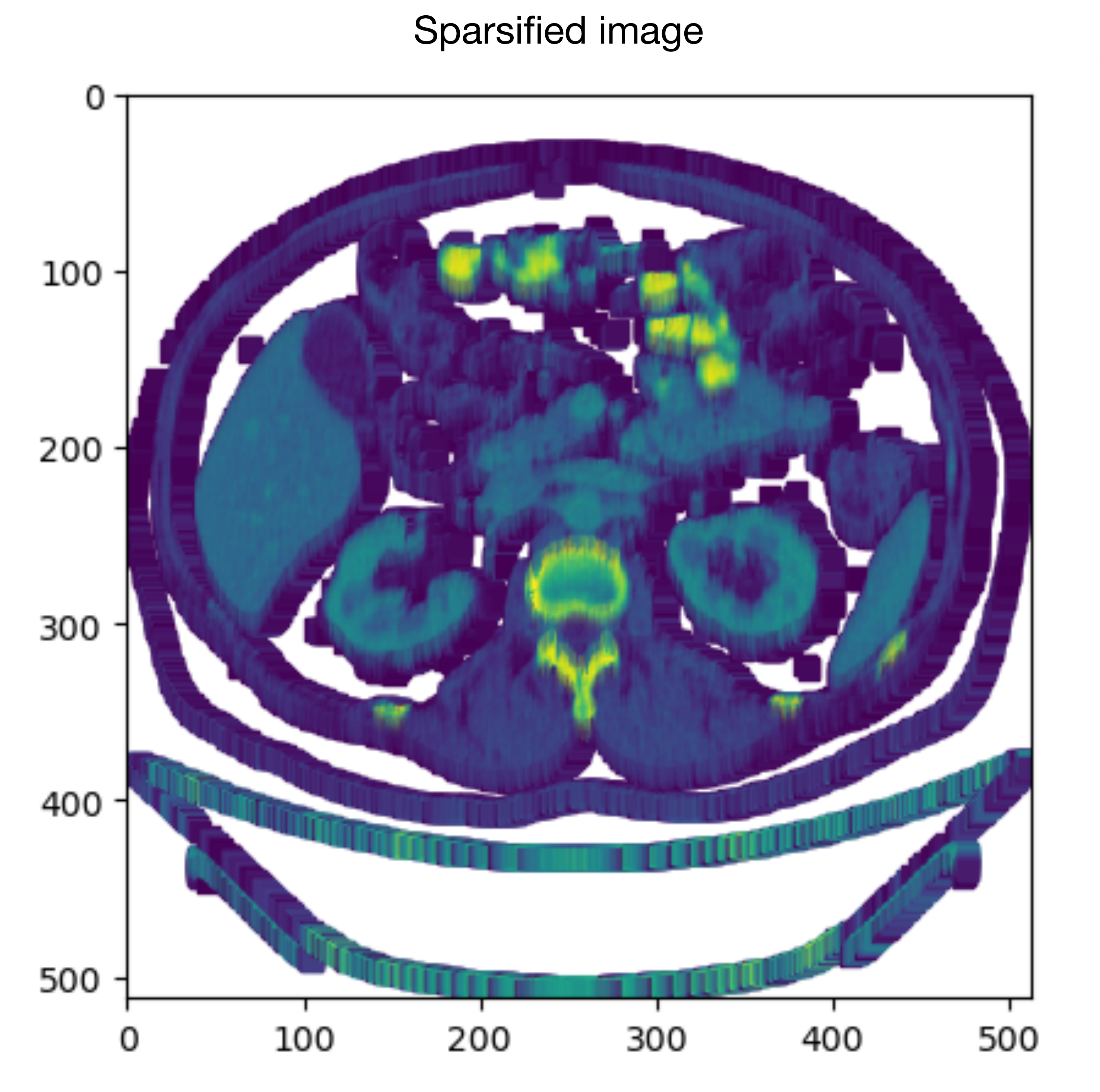}\includegraphics[width=0.33\textwidth,height=0.3\textwidth]{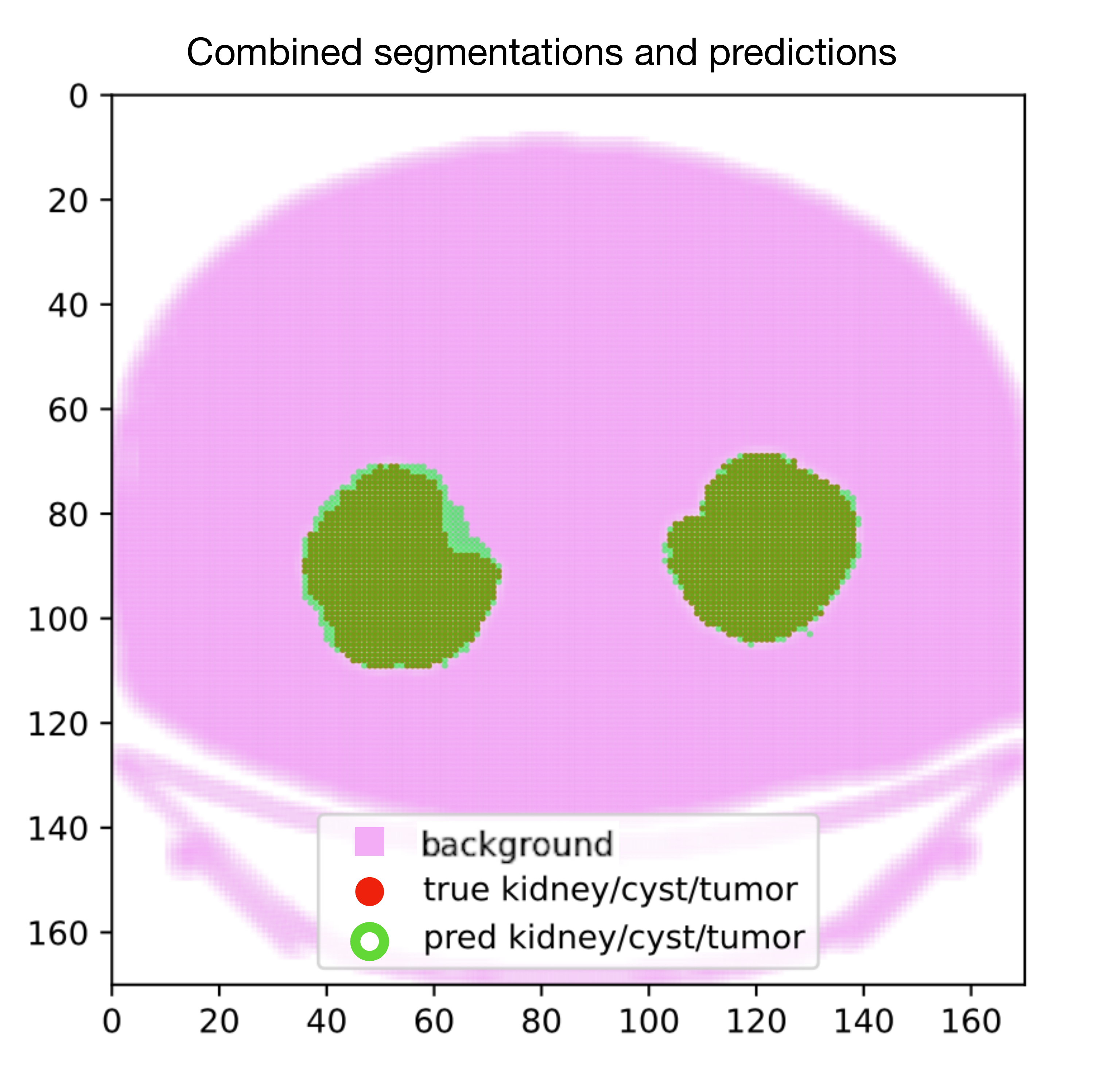}\\
     \includegraphics[width=0.33\textwidth,height=0.3\textwidth]{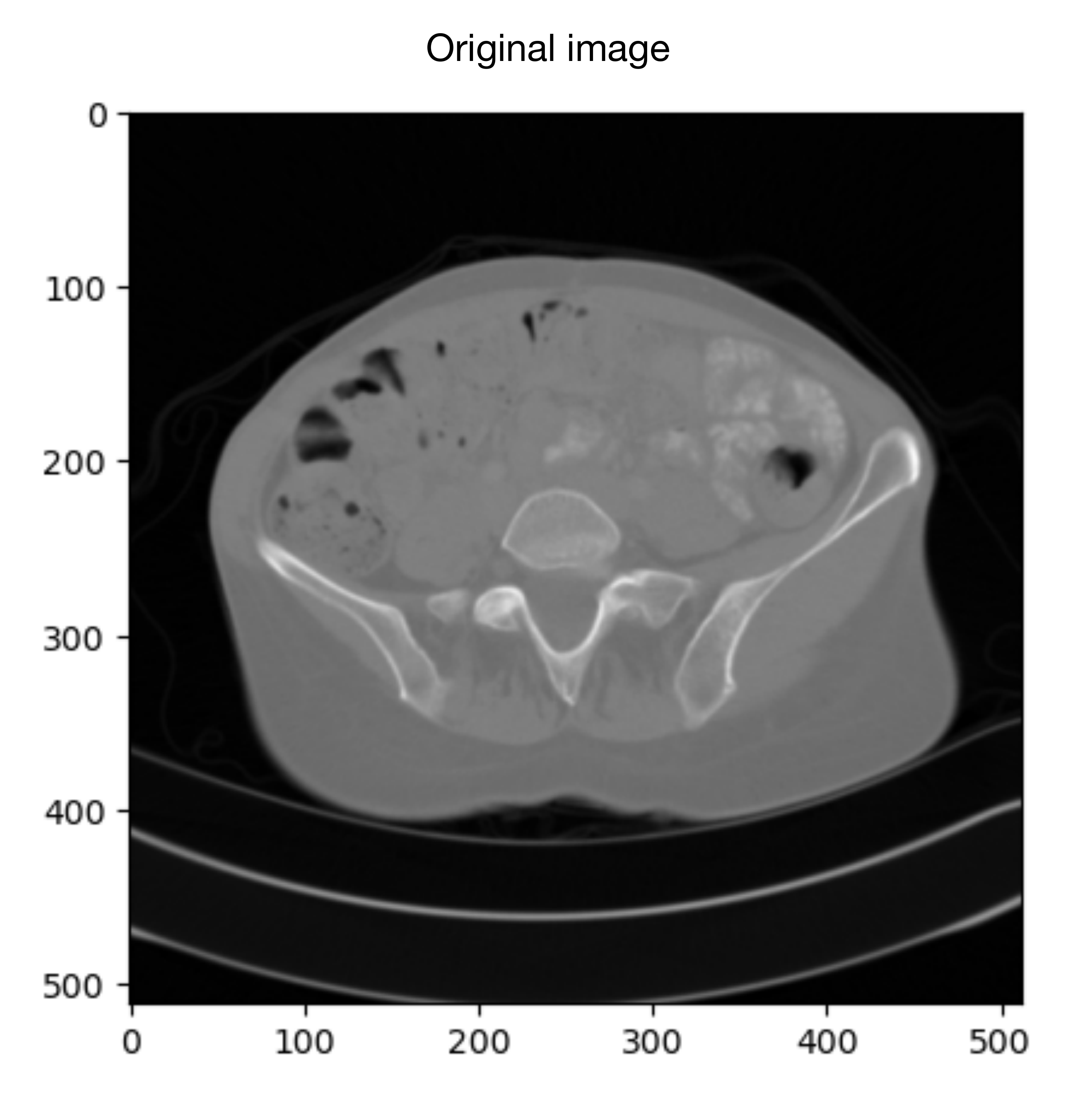}\includegraphics[width=0.33\textwidth,height=0.3\textwidth]{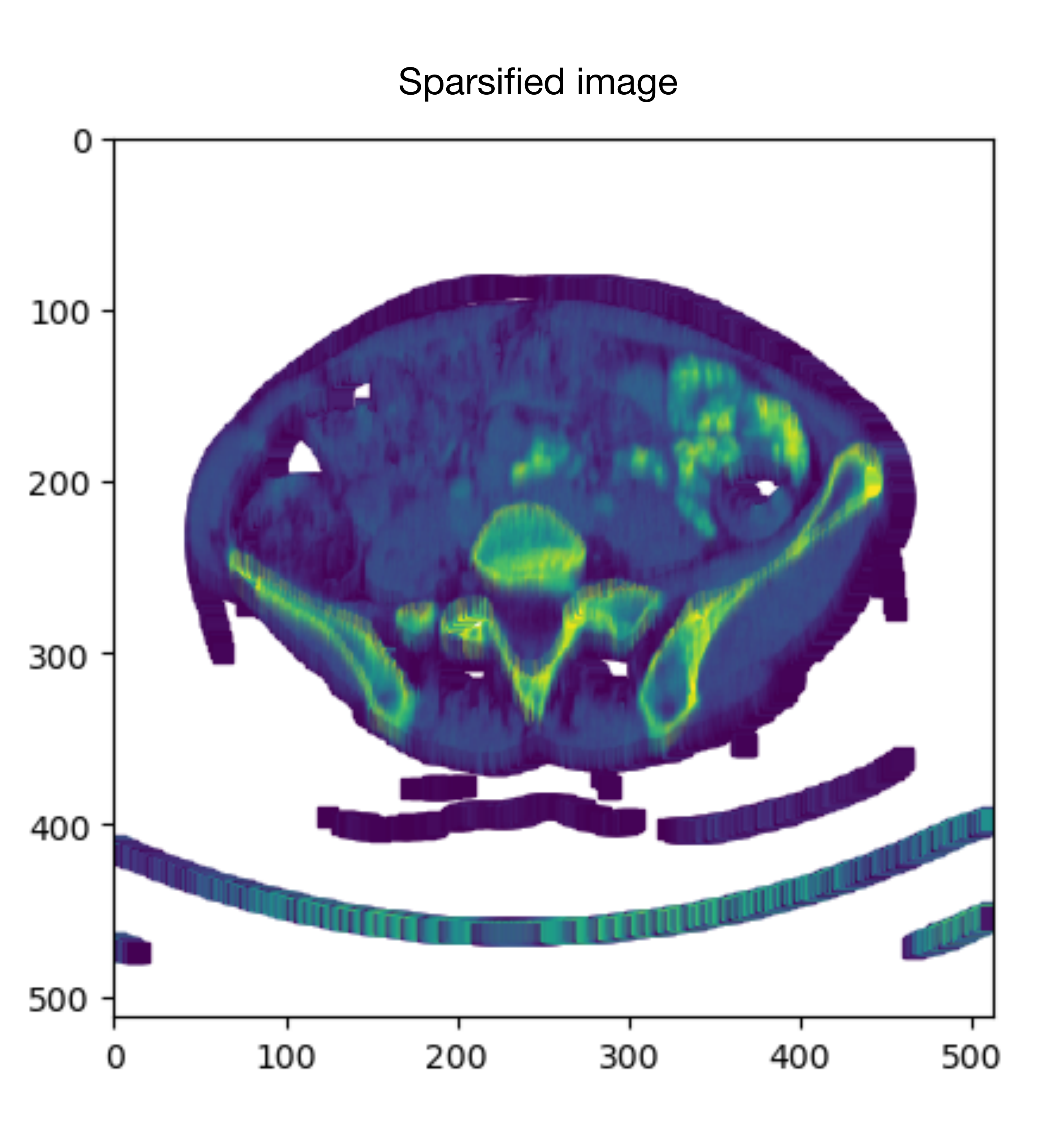}\includegraphics[width=0.33\textwidth,height=0.3\textwidth]{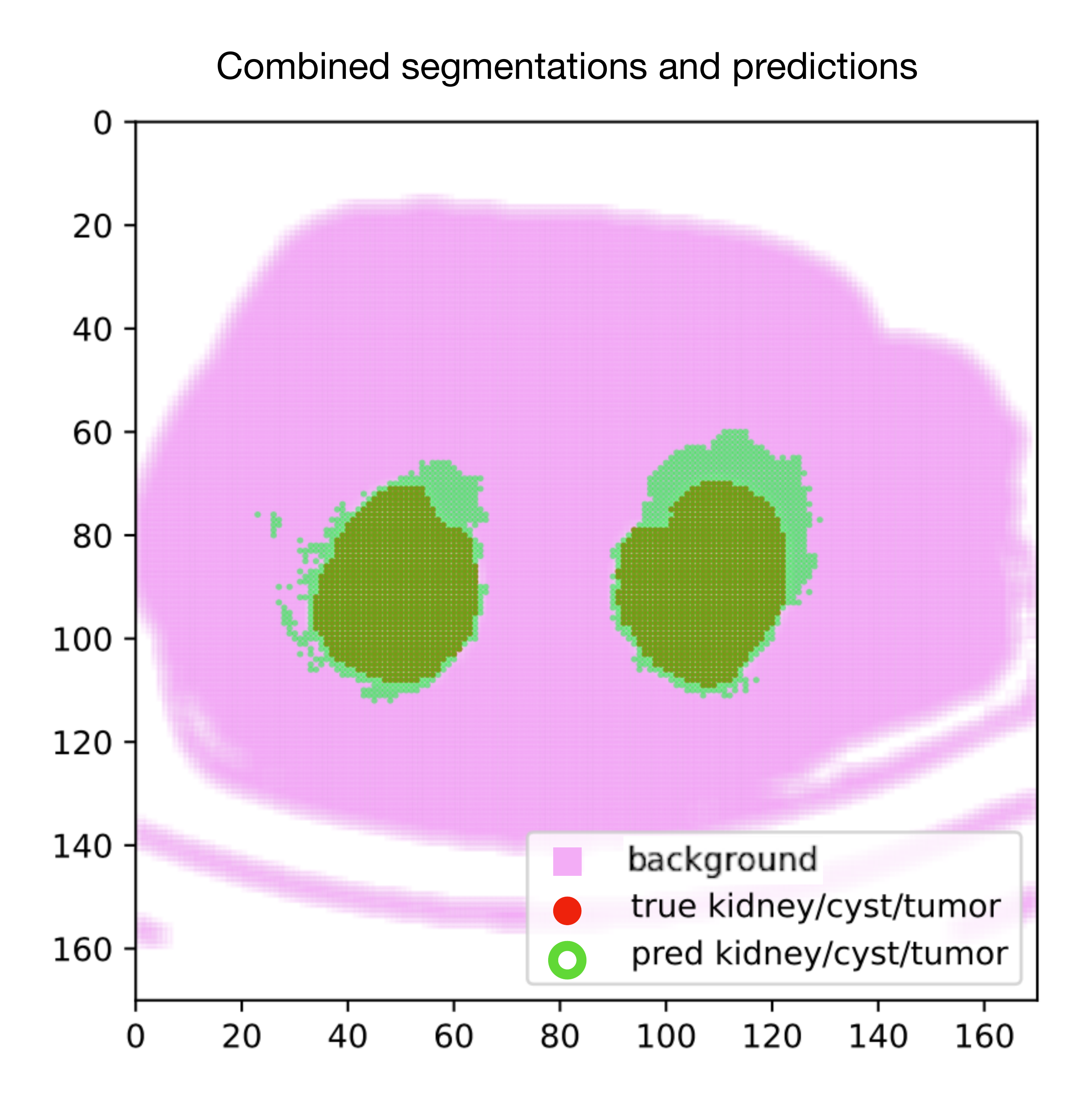}\\
      \includegraphics[width=0.33\textwidth,height=0.3\textwidth]{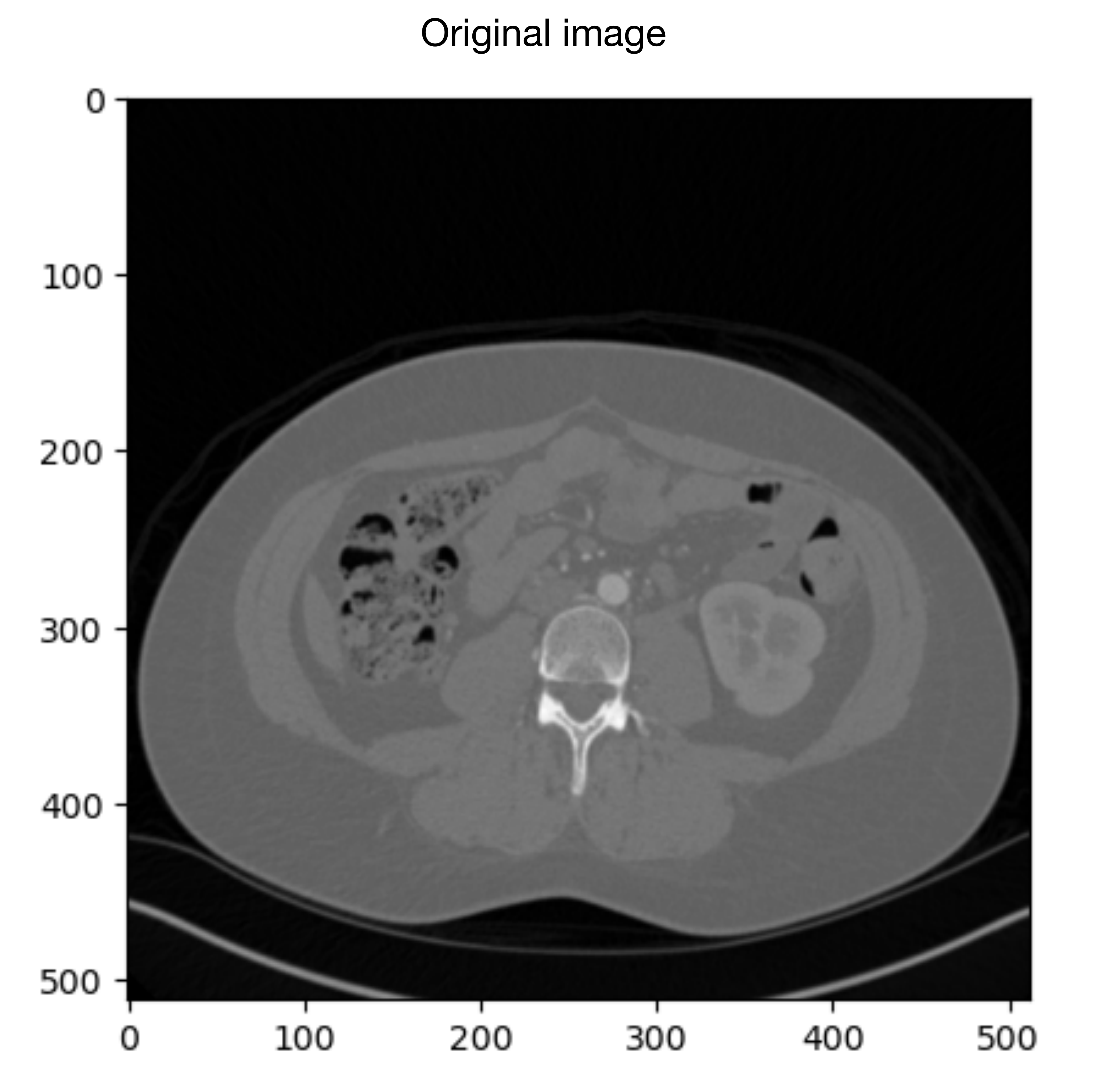}\includegraphics[width=0.33\textwidth,height=0.3\textwidth]{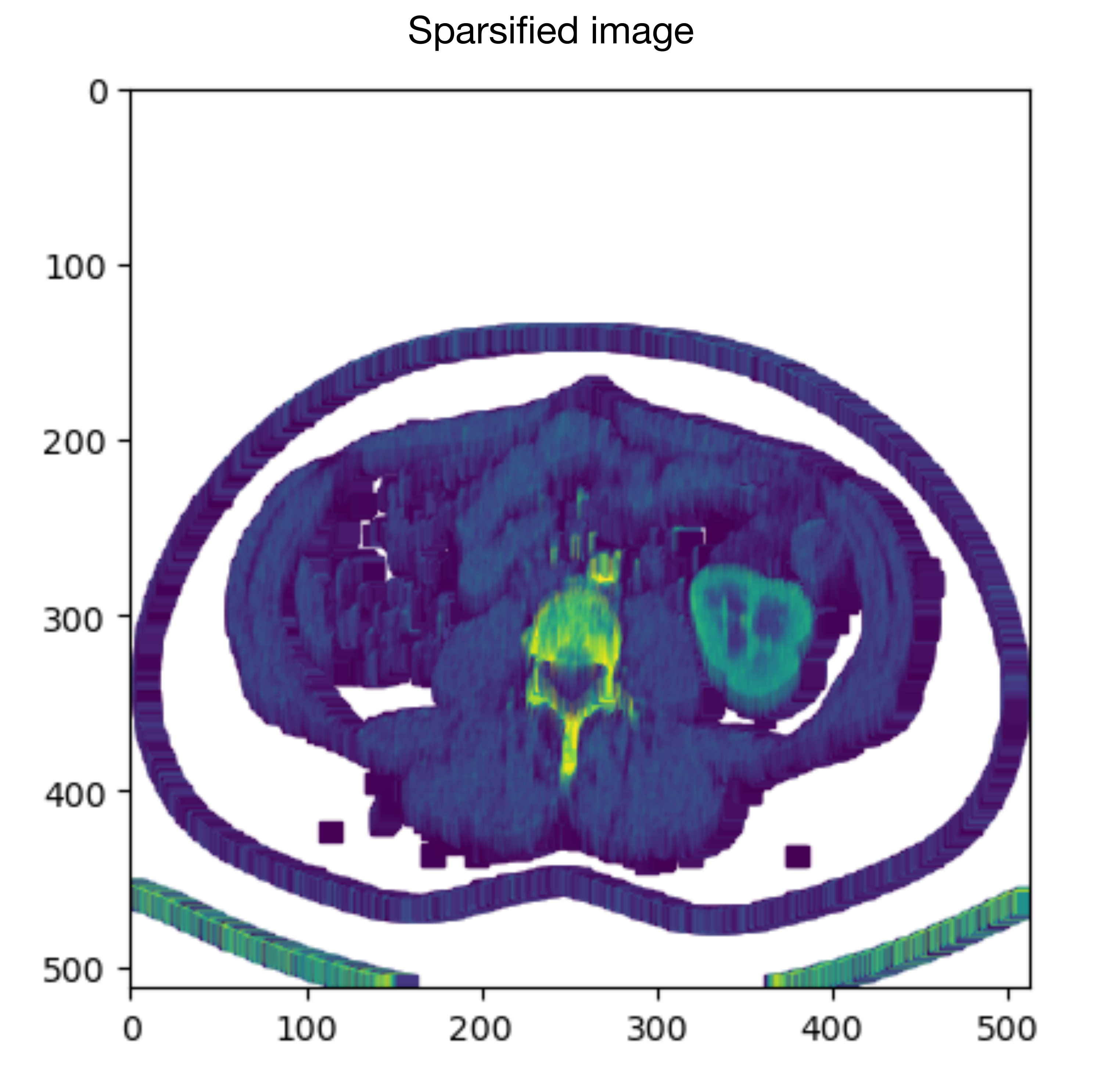}\includegraphics[width=0.33\textwidth,height=0.3\textwidth]{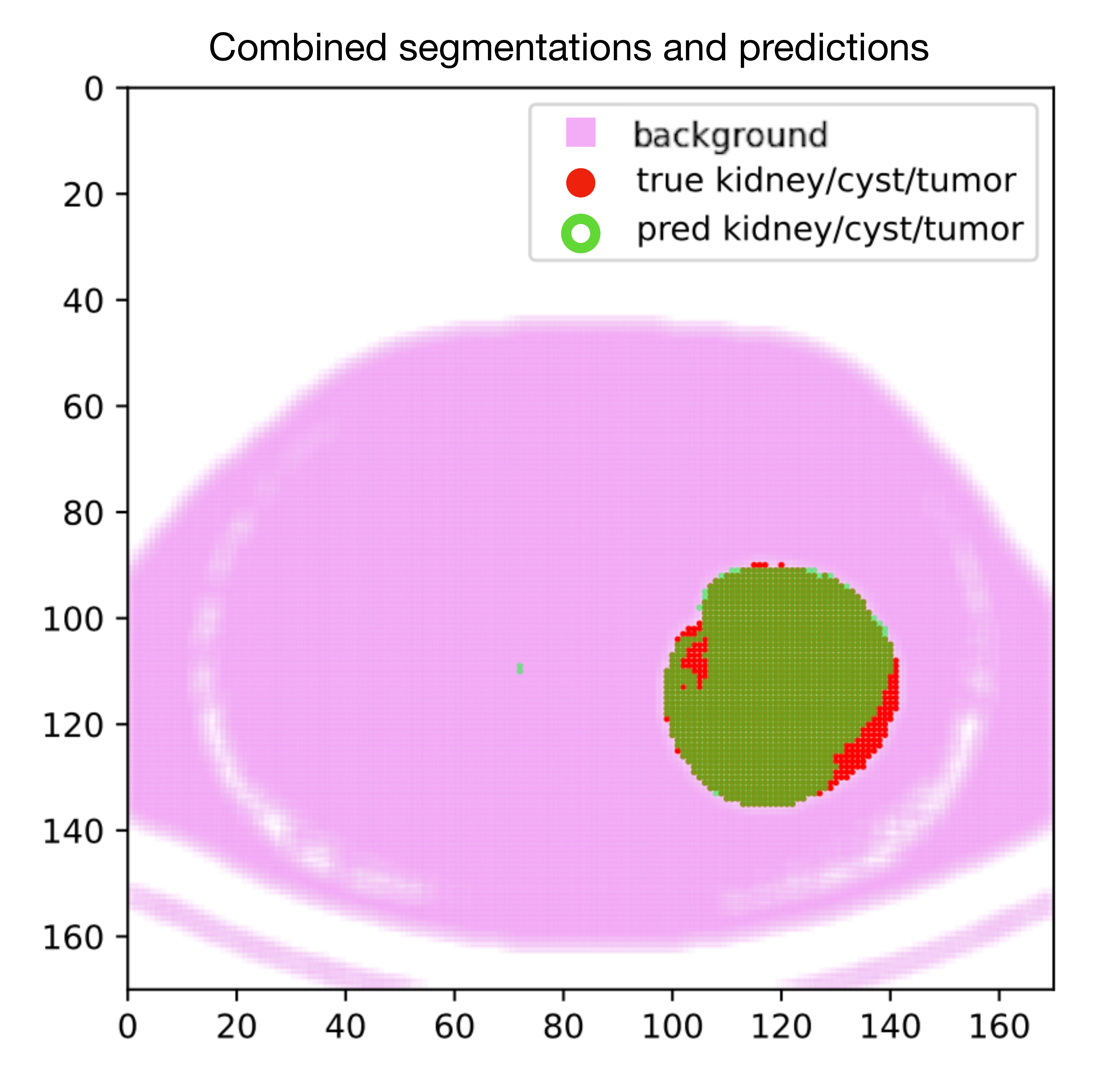}
  \caption{From left to right: (1) central 2D slice from the original scan; (2) corresponding sparsified image input to the SSCN; (3) combination of all voxelised 2D slices of the scan, showing background, true and predicted segmentations using the trained SSCN. Each row is a different case (patient), and these examples have been chosen as illustrations of good overlap between true and predicted segmentations, and the prediction of a single kidney in a case in which only one kidney was present (bottom).}
  \label{fig:results_good}
\end{figure*}

\begin{figure*}[th]
     \includegraphics[width=0.33\textwidth,height=0.3\textwidth]{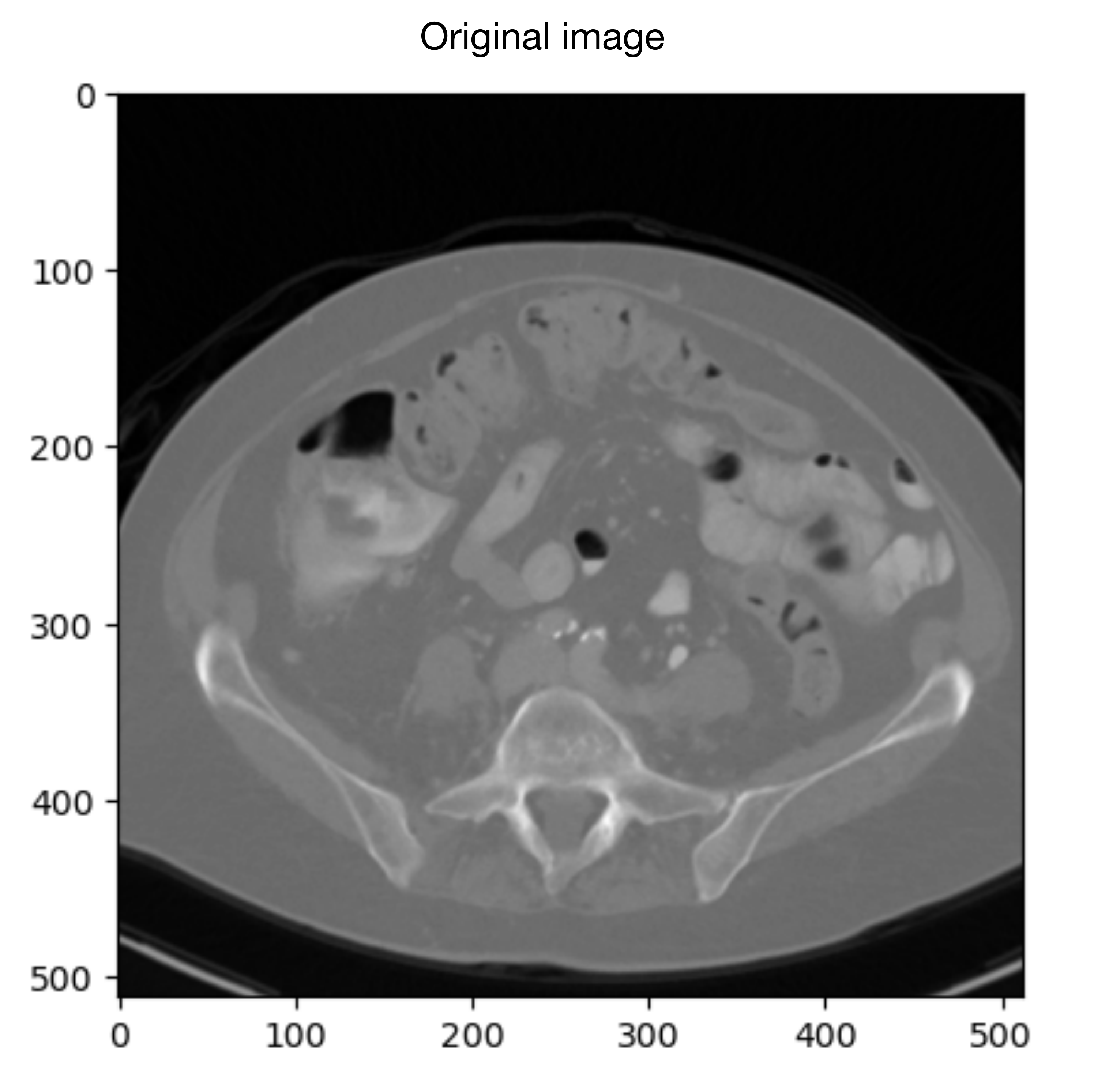}\includegraphics[width=0.33\textwidth,height=0.3\textwidth]{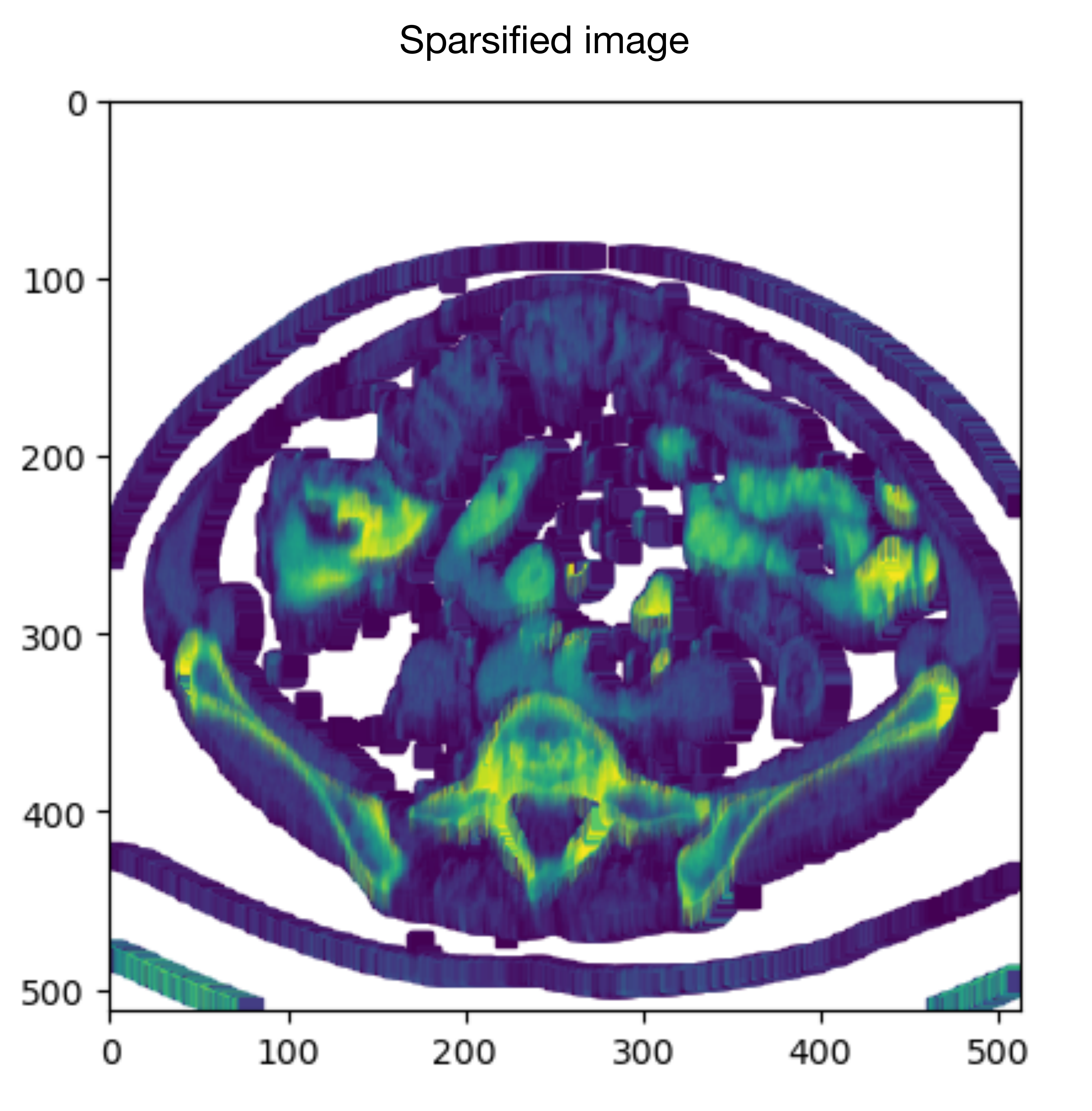}\includegraphics[width=0.33\textwidth,height=0.3\textwidth]{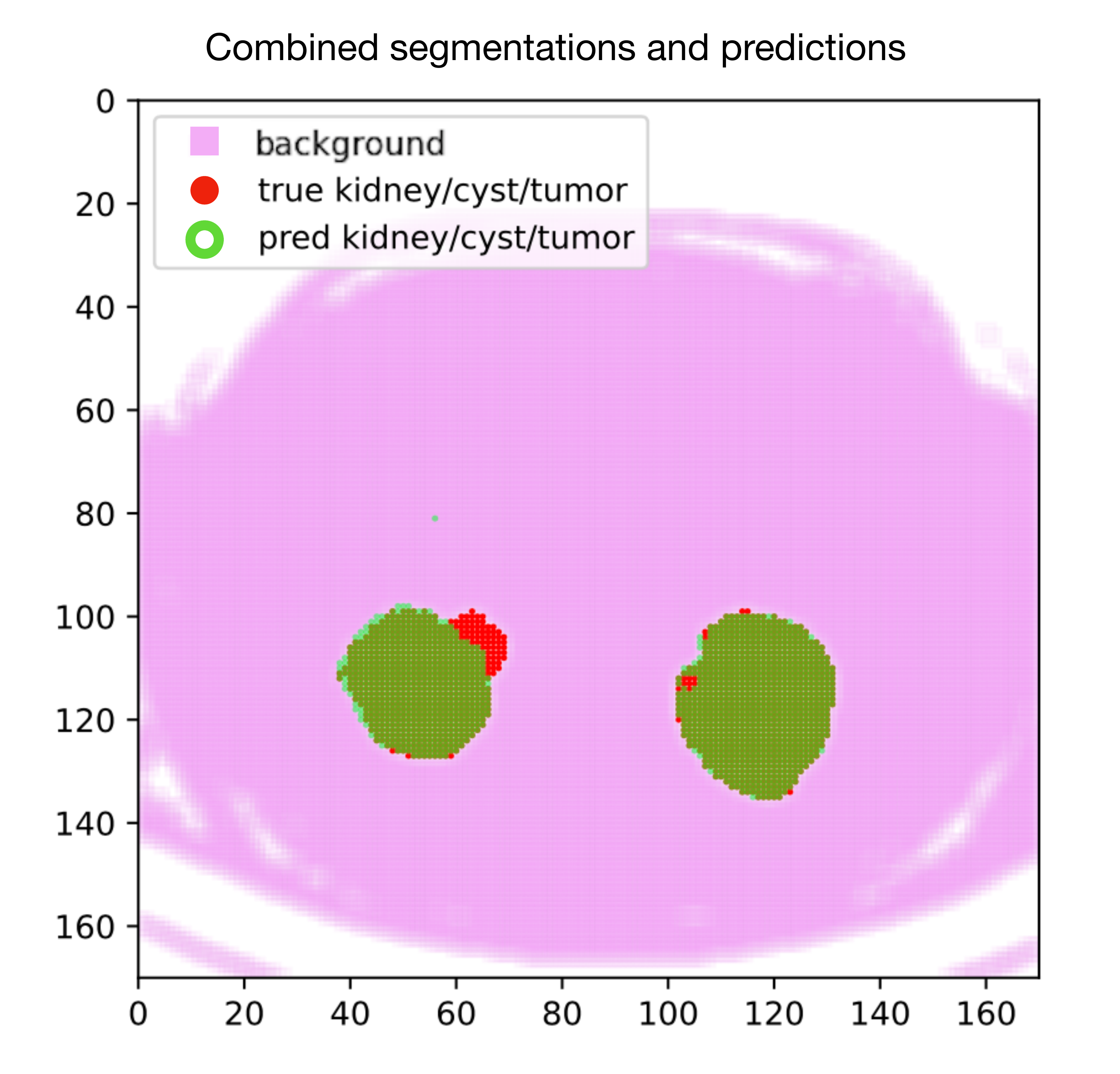}\\
     \includegraphics[width=0.33\textwidth,height=0.3\textwidth]{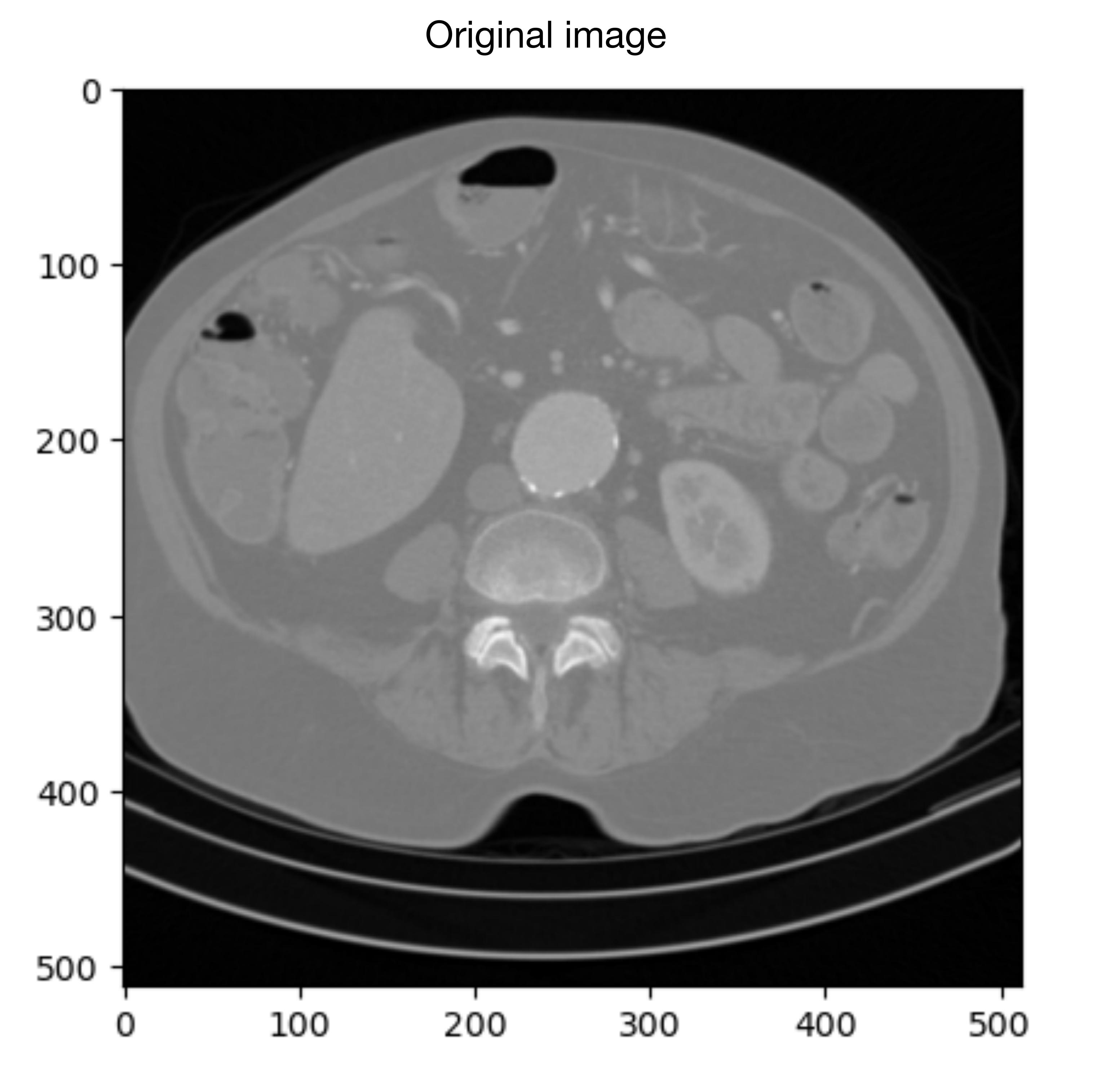}\includegraphics[width=0.33\textwidth,height=0.3\textwidth]{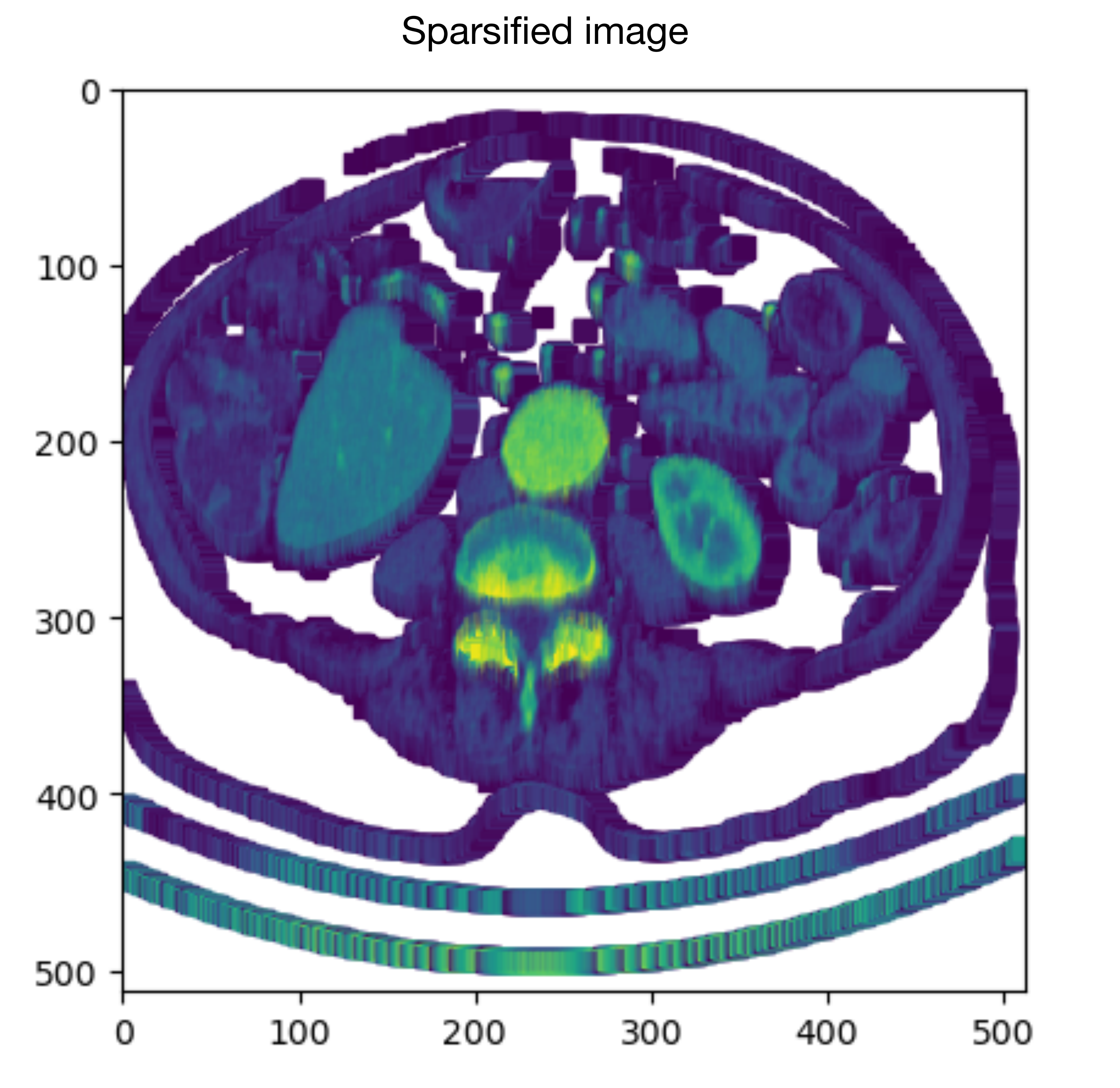}\includegraphics[width=0.33\textwidth,height=0.3\textwidth]{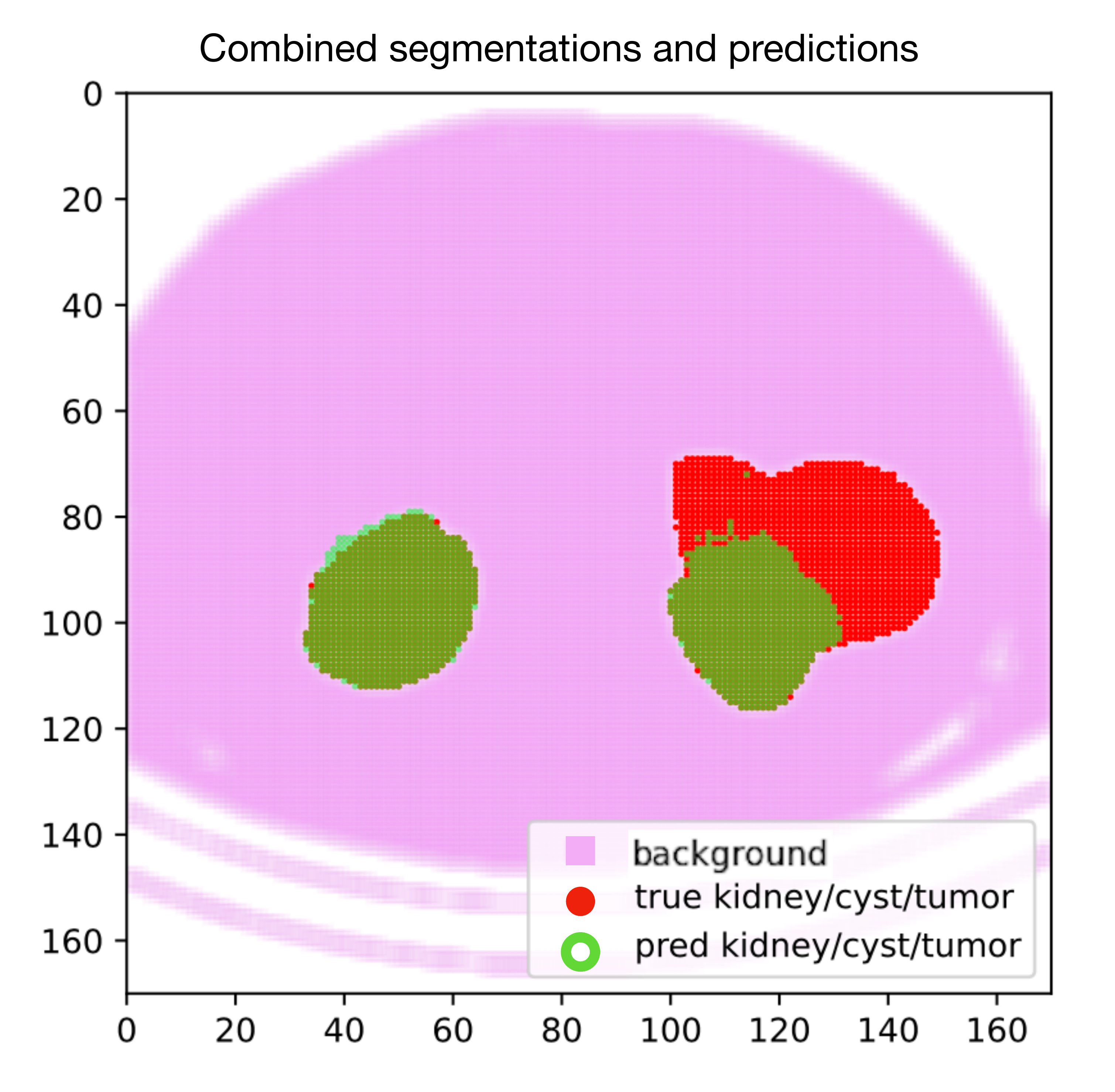}\\
      \includegraphics[width=0.33\textwidth,height=0.3\textwidth]{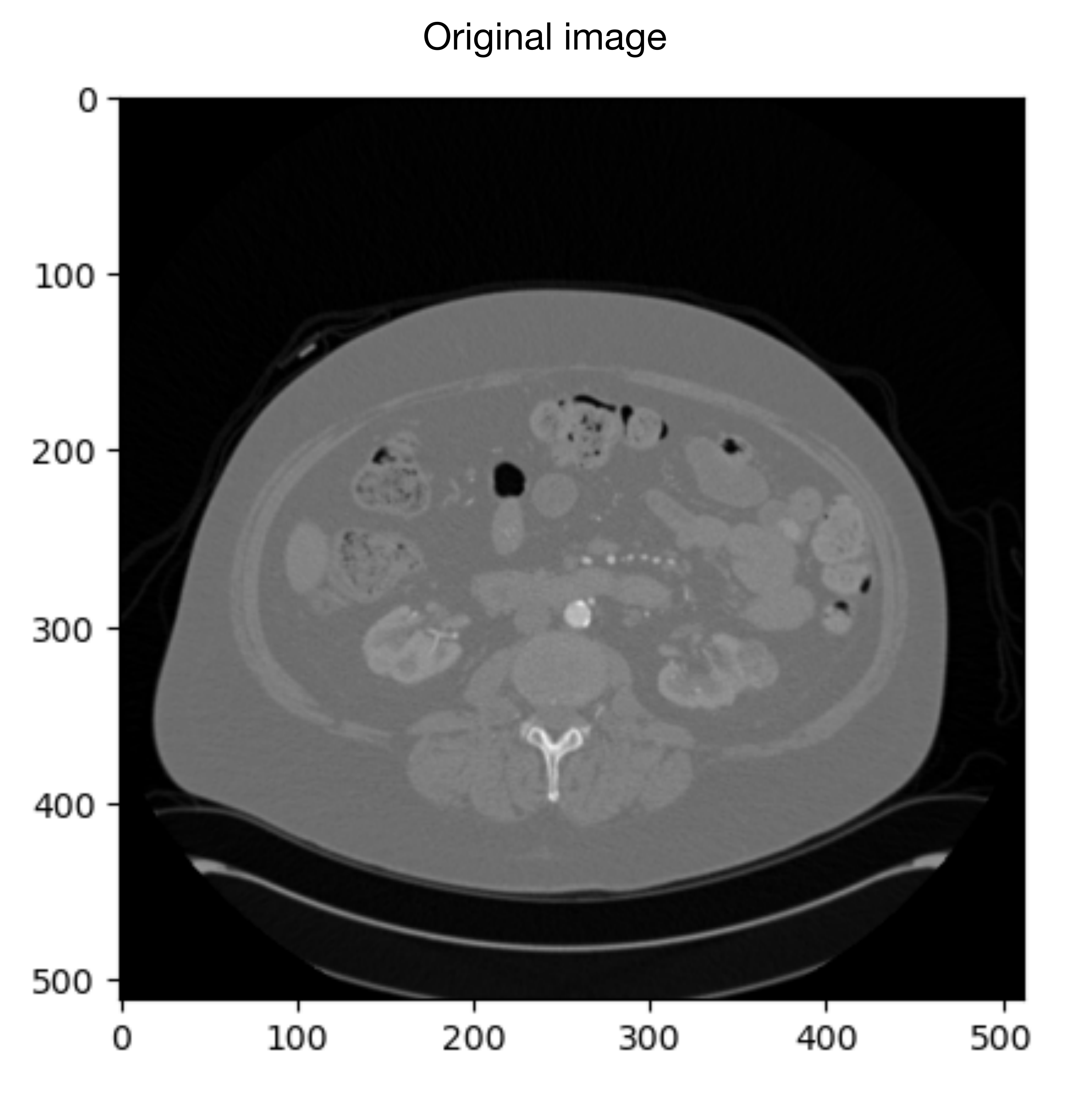}\includegraphics[width=0.33\textwidth,height=0.3\textwidth]{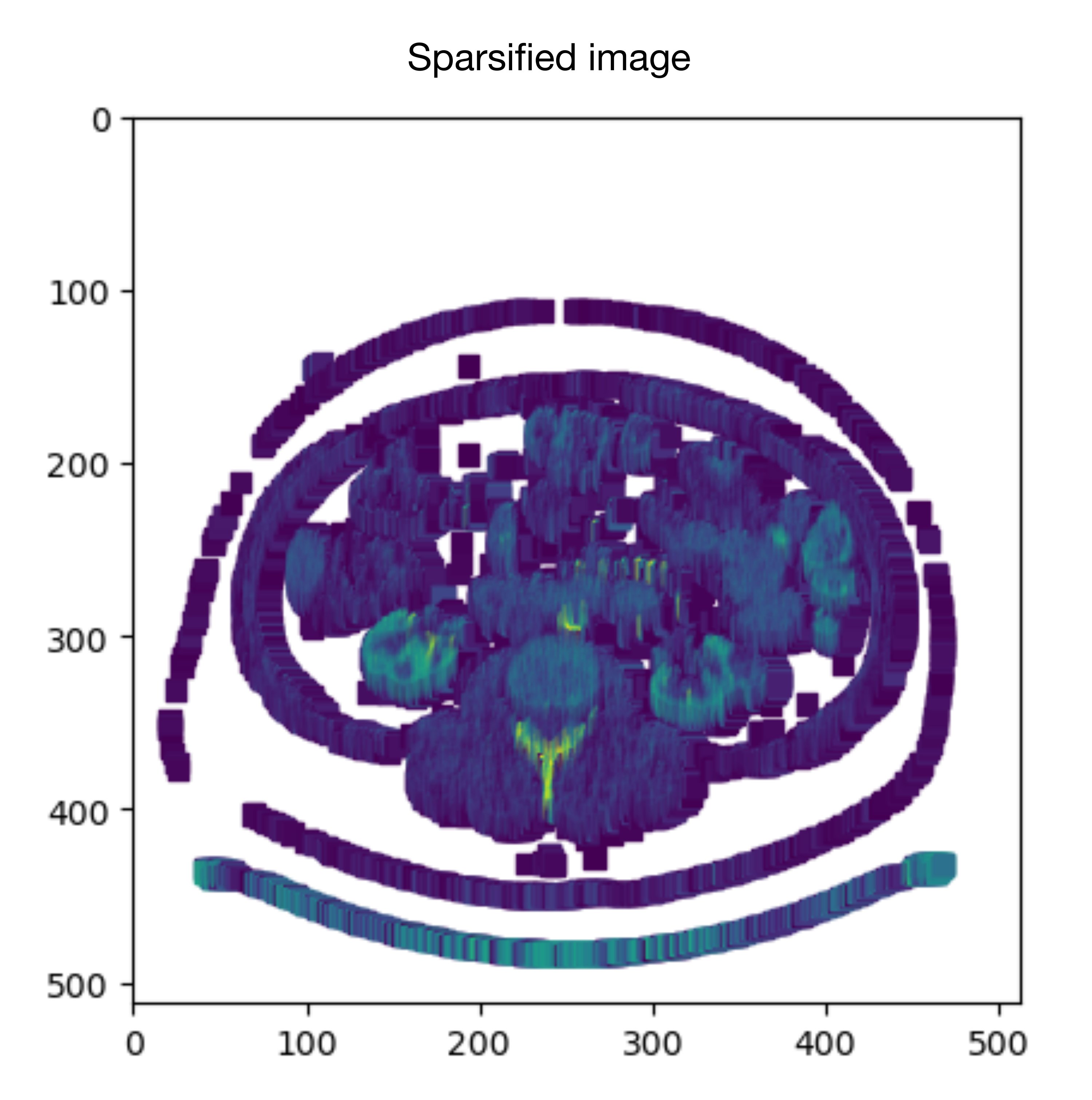}\includegraphics[width=0.33\textwidth,height=0.3\textwidth]{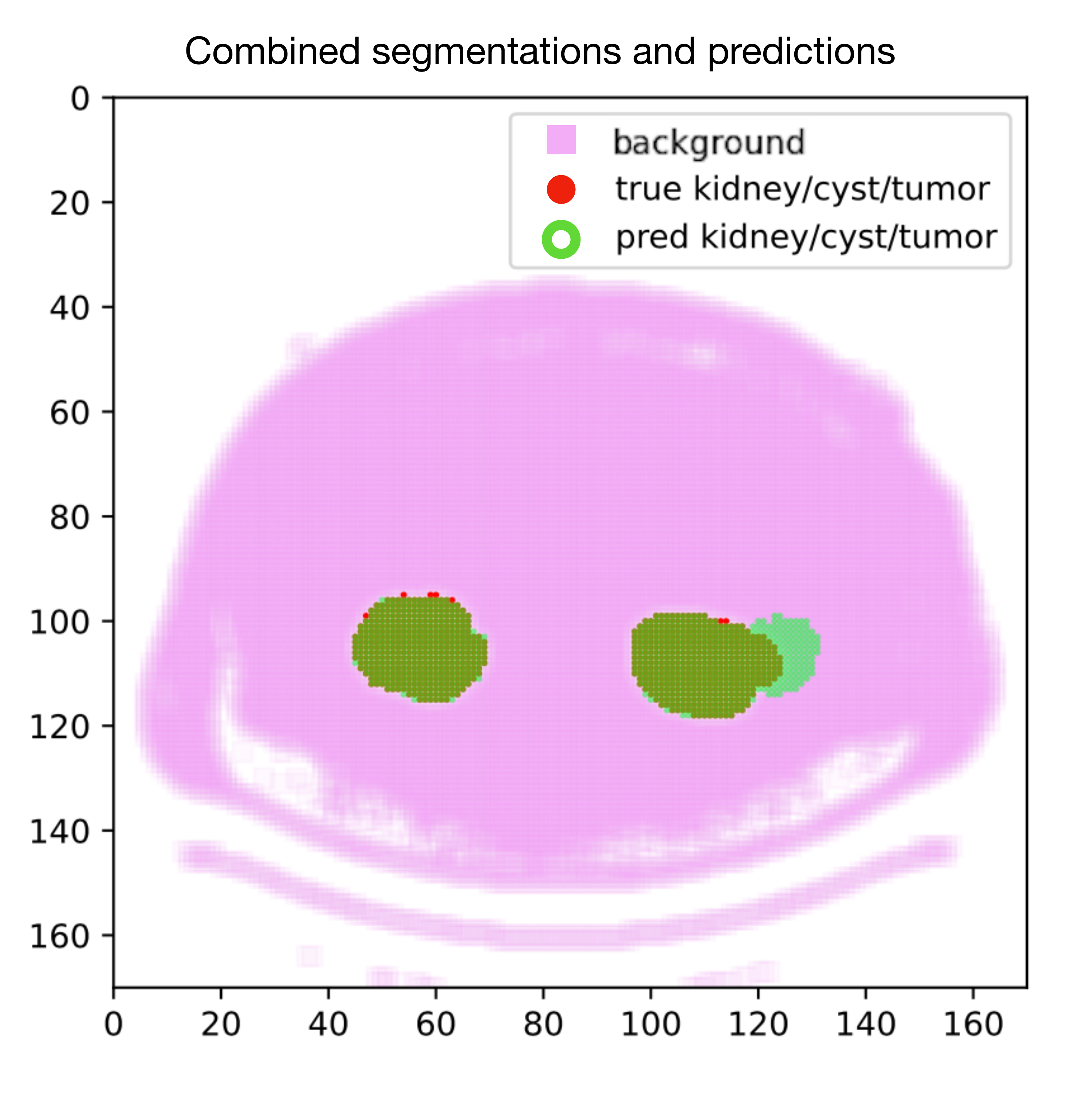}\\
       \includegraphics[width=0.33\textwidth,height=0.3\textwidth]{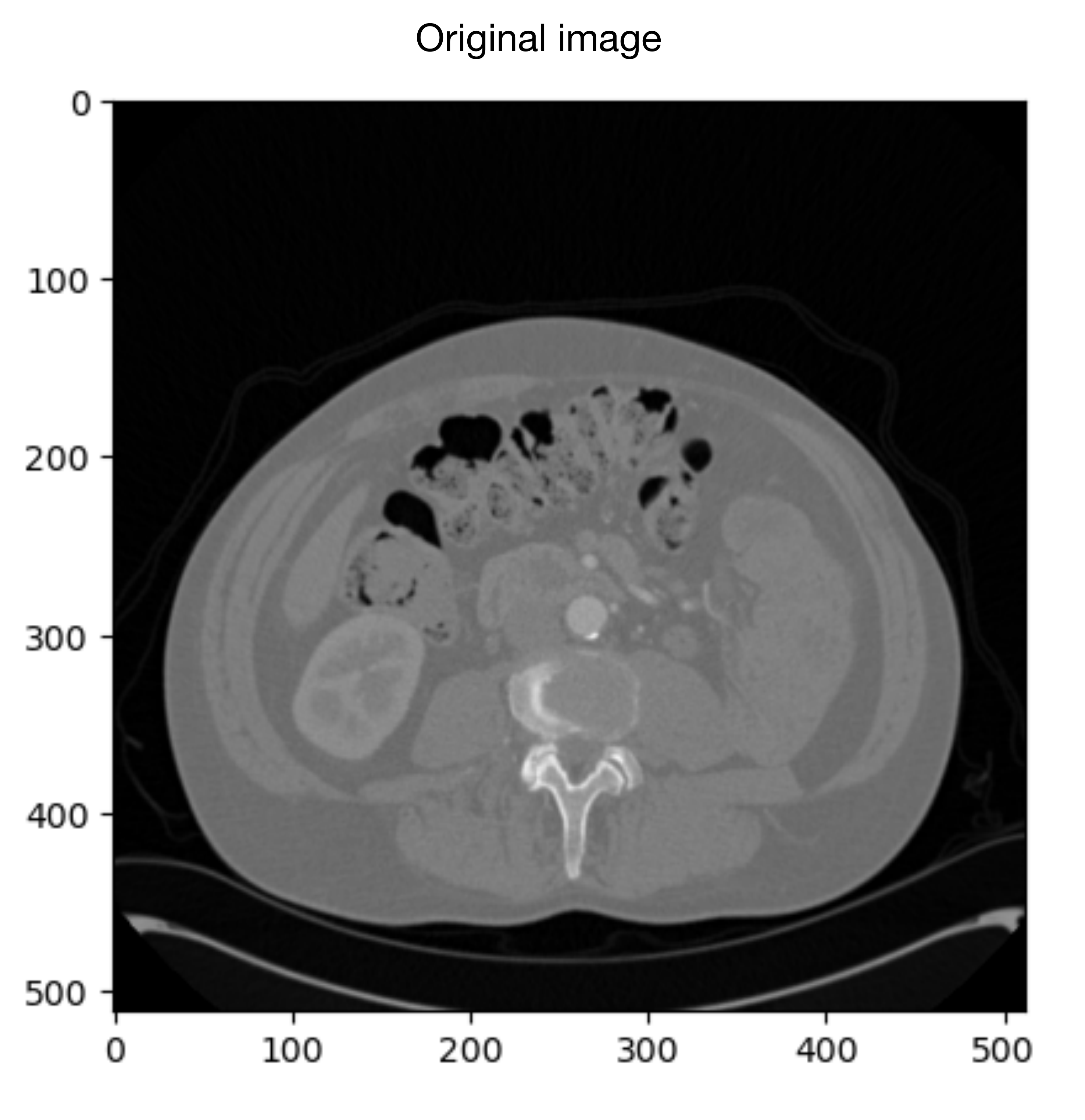}\includegraphics[width=0.33\textwidth,height=0.3\textwidth]{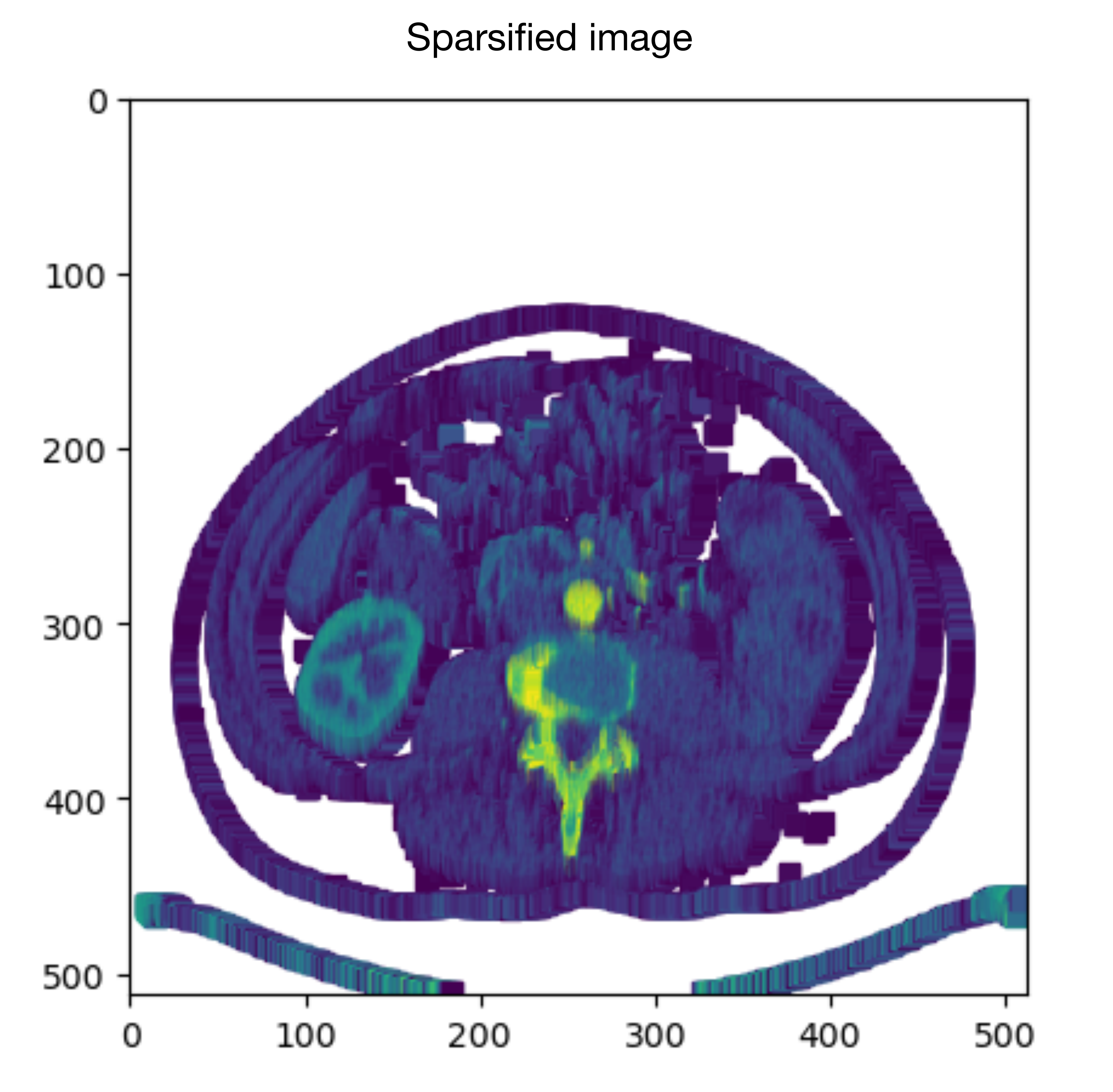}\includegraphics[width=0.33\textwidth,height=0.3\textwidth]{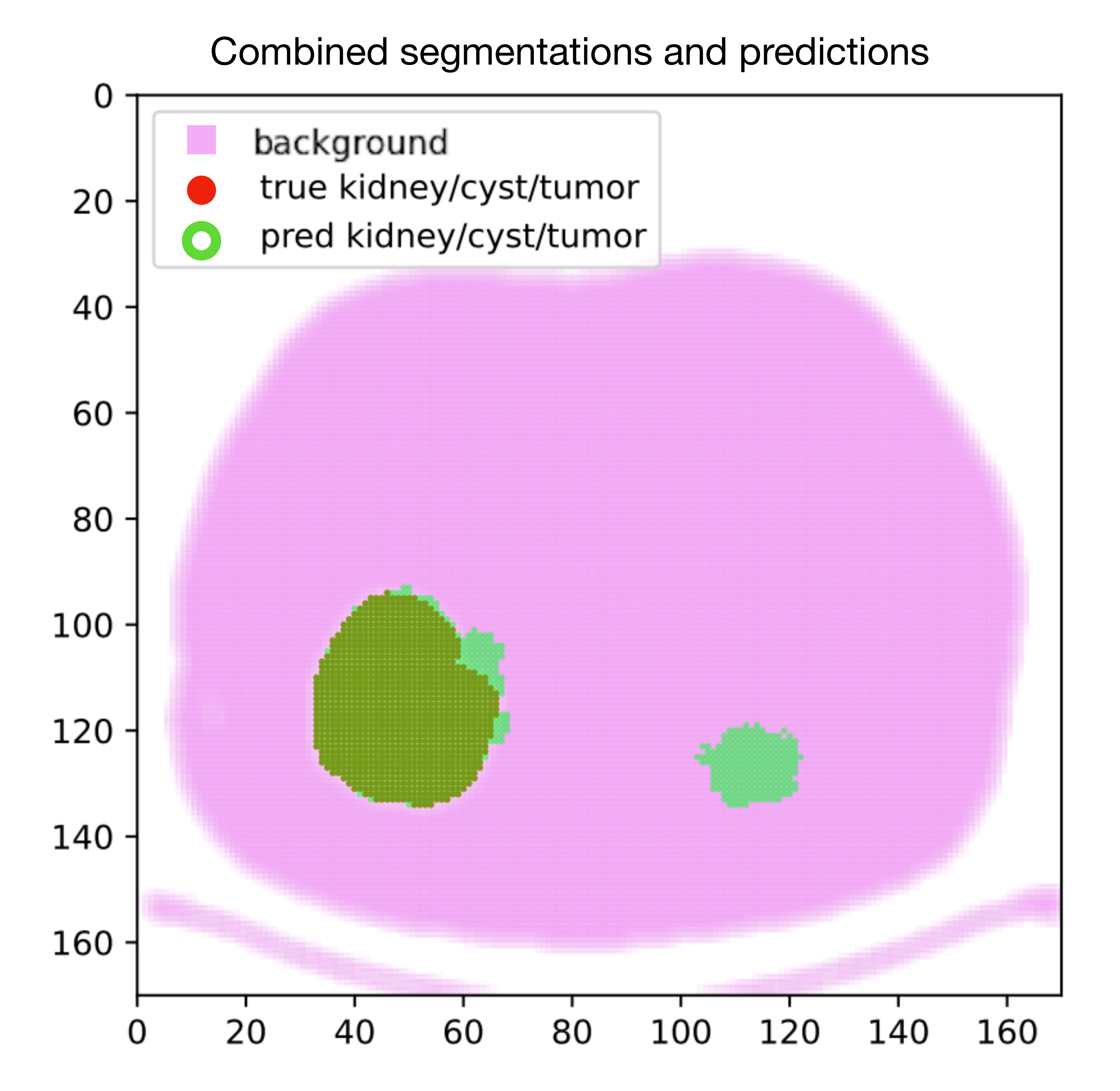}
  \caption{From left to right: (1) central 2D slice from the original scan; (2) corresponding sparsified image input to the SSCN; (3) combination of all 2D slices of the scan, showing background, true and predicted segmentations using the trained SSCN. Each row is a different case (patient), and these examples have been chosen as illustrations of imperfect overlap between true and predicted segmentations, and a small prediction of a second kidney in a case in which only one kidney was present (bottom).}
  \label{fig:results_soso}
\end{figure*}

\subsection{Sparsification} 
The process of sparsification requires to ignore certain voxels from the input images, those deemed unimportant to the segmentation task. Different ranges can be chosen for the problem, with a variable fraction of voxels discarded for both background and foreground (kidney + masses). We optimised this range by taking into account the maximum number allowed by a generic GPU card (for example, the memory of an NVIDIA V100 GPU can store a maximum of $\sim$1M pixels/voxels). Considering different 3D downsampling factors (1,2,3,4,5), we set the maximum number of pixels allowed to (1, 8, 27, 64, 125)M, respectively, and found a range of Hounsfield units for each one that minimises the signal loss. The different ranges can be found in Table~\ref{tab:factors}. 

For the remainder of this study, the range for intensities [-30, 350] H.U. was used, motivated by the results in Table~\ref{tab:factors}. The position of the minimum and maximum values in the range is shown visually in Fig.~\ref{fig:min_max_pixels}, showing the cumulative fraction of pixels removed by this minimum (maximum) threshold in the top (bottom) panel. The green arrows show the range chosen, demonstrating the good retention of kidney and mass voxels (red, $\sim$\,98\% retained) while rejecting most background voxels (blue, $\sim$\,77\% removed). In total, this sparsification process reduces the number of voxels, without any downsampling, by a factor of two per CT scan.
 
\subsection{Region of interest (ROI) finder} 
The next stage after the sparsification process is the region of interest (ROI) finder, for which the Minkowski Engine v0.5.4 framework was used. The Minkowski Engine~\cite{choy20194d} is an auto-differentiation library that supports generalised sparse convolutions, designed to deal with sparse input images.

Each input image was quantised by a factor of 3 (in each dimension) so that the full 3D stack can fit into GPU memory for any of the cases (ranging from 29 to 1059 2D slices). The architecture of the neural network chosen was a sparse submanifold version of the well-known UNet-34~\cite{unet}, originally proposed for biomedical image semantic segmentation. The goal of the network is to learn to classify each voxel of the input images as either ``kidney/cyst/tumour'' (signal) or ``nothing'' (background).

We used Python version 3.10.4 and PyTorch version 1.11.0. For the ROI finder network, Adam was our choice for the optimiser, achieving the best results with a learning rate of $10^{-4}$, and a batch size of 2\footnote{We can only fit two images per batch due to their large size.}. The final tests were run on an NVIDIA A100 GPU; however, initial tests were done using NVIDIA V100 GPU with comparable results.

Due to the low number of images available for training, we performed a 10-fold cross-validation procedure as follows: the dataset was shuffled and then divided into ten equal parts, where eight were used for training, one for validation, and one for testing; this procedure is repeated ten times, each time employing a different part for validation and testing. In each step of the 10-fold cross-validation, a new network instance is trained independently on the eight corresponding remaining parts. 

\subsection{Full segmentation} 
The output of the first stage (ROI finder) will be used in a second stage to crop regions of the images in 3D where signal (kidney + cyst + tumour) is likely to exist. To do this, we will use limits in (x,y) per 2D slice that define a rectangle enclosing both kidneys (i.e. treating all signal in the slice as one), discarding voxels outside, and a z range for the entire 3D volume, discarding whole slices where no signal appears to be present. In this way, the original 3D scans will be reduced in size, then input to an SSCN without quantisation (full resolution). 

\section{Results} 
\label{sec:results}
Examples of the sparsification process, applying the range of [-30, 350] H.U. to the input images and true segmentations, are presented in Figs.~\ref{fig:results_good} and~\ref{fig:results_soso} (left and central columns). In these images, the central 2D slice of each 3D volume was selected per row (case), and the voxels removed during the process of sparsification are shown in white. 

The rightmost column of Figs.~\ref{fig:results_good} and~\ref{fig:results_soso} shows the resulting ROIs predicted by the trained SSCN, combining the result of all 2D slices into a single image, overlapping the true segmentations (red) and the predicted ROIs (green). The effect of the voxelisation in the first stage (ROI finder) is also visible in these panels. 

Visually, these ROIs show in general a good agreement with the truth, as it can be observed in Fig.~\ref{fig:results_good}. In Fig.~\ref{fig:results_soso} we present examples of some of the limitations we have observed in particular cases. It is interesting to highlight that the method is able to predict a single ROI when only one kidney is present (Fig.~\ref{fig:results_good}, bottom row), or to create only a small ROI (Fig.~\ref{fig:results_soso}, bottom row).

The Dice similarity coefficient (DSC) obtained for signal (kidneys/cysts/tumours) for all images in this first stage was 84.6\%\footnote{Note this is still voxelised by a factor of 3.}, while the training took only 2-3 minutes per epoch. The results of the training of this ROI finder stage can be found in Table~\ref{tab:ROI_training}.

\begin{table}
\centering
\caption{Results of the training of the first stage (ROI finding), combining the testing results from the 10 independent networks. The accuracy obtained was 99.1\%.}
\vspace{5pt}
\resizebox{1.0\linewidth}{!}{
\begin{tabular}{r | c | c | c | c }
     & \textbf{precision} & \textbf{recall} & \textbf{f1-score} & \textbf{support} \\
     \hline
     \rule{0pt}{2ex} Background & 0.995  &  0.995  &  0.995 & 176634333\\
       \hline
    \rule{0pt}{2ex} Signal & 0.845   &  0.847  &   0.846  & 5600613\\
     \hline
\end{tabular}}
\label{tab:ROI_training}
\end{table}

The ROI found per scan will be used to crop the original images for the input for the second stage. It is interesting to quantitatively analyse the ROI-based cropping of the images from two perspectives: (1) by calculating how much signal (background) is kept (removed) when applying a 2D cut on all the slices of the CT scan; and (2) by calculating how many slices without actual signal are effectively removed. Table~\ref{tab:roi_cropping} shows the cropping performance in different scenarios where we added a safety margin of a few voxels/slices to the (x,y,z) limits calculated for each case. As expected, the larger the cropping window, the more signal is kept, but the less background is removed. 

\begin{table}
\centering
\caption{Effectiveness of the ROI finding for cropping the images, combining the testing results from the 10 independent cross-validation networks. The table shows the average amount of signal kept and background removed for the 2D projected voxels of the CT scan and the slices. The first column indicates how much the cropping window is enlarged for each specific test.}
\vspace{5pt}
\resizebox{1.0\linewidth}{!}{
\begin{tabular}{c | c | c | c | c }
     & \multicolumn{2}{c|}{\textbf{2D projected voxels}} & \multicolumn{2}{c}{\textbf{slices}} \\
     \hline
     \multirowcell{2}{\textbf{\shortstack{$\pm$ voxels/\\slices}}} & \multirowcell{2}{\textbf{\shortstack{signal\\(kept)}}} & \multirowcell{2}{\textbf{\shortstack{background\\(removed)}}} & \multirowcell{2}{\textbf{\shortstack{signal\\(kept)}}} & \multirowcell{2}{\textbf{\shortstack{background\\(removed)}}} \\
     & & & & \\
     \hline
    \rule{0pt}{2ex} 0 & 0.950 & 0.900 & 0.977 & 0.867 \\
    \rule{0pt}{2ex} 1 & 0.959 & 0.890 & 0.984 & 0.822 \\
    \rule{0pt}{2ex} 2 & 0.964 & 0.880 & 0.987 & 0.773 \\
    \rule{0pt}{2ex} 3 & 0.968 & 0.869 & 0.989 & 0.724 \\
    \rule{0pt}{2ex} 4 & 0.971 & 0.857 & 0.991 & 0.675 \\
    \rule{0pt}{2ex} 5 & 0.974 & 0.846 & 0.992 & 0.626 \\
    \rule{0pt}{2ex} 6 & 0.976 & 0.834 & 0.993 & 0.576 \\
    \rule{0pt}{2ex} 7 & 0.979 & 0.821 & 0.994 & 0.526 \\
    \rule{0pt}{2ex} 8 & 0.981 & 0.809 & 0.995 & 0.476 \\
    \rule{0pt}{2ex} 9 & 0.982 & 0.796 & 0.996 & 0.427 \\
    \rule{0pt}{2ex} 10 & 0.984 & 0.782 & 0.997 & 0.377 \\
     \hline
\end{tabular}}
\label{tab:roi_cropping}
\end{table}

\section{Discussion}
In this paper, we demonstrated the feasibility and potential for using submanifold sparse convolutional networks (SSCN) for the segmentation of organs and tumours in CT images. To our knowledge this is the first time that such networks have been applied to the segmentation of CT images, and only one more example exists in the literature in which the Minkowski engine has been applied to medical imaging at all~\cite{Li2022}.

We proposed a methodology that deploys SSCNs in three steps, starting with a process of sparsification applied to the original images. The ranges for sparsification based on H.U. were optimised accordingly to the ranges observed for signal in the dataset used, taking into account the maximum number of voxels that can be stored in GPU memory. 

After the sparsification step, a first SSCN was trained to perform region of interest (ROI) selection, which obtains a good accuracy of $\sim$\,99.1\%, with a precision of $\sim$\,84.5\% ($\sim$\,99.5\%) to select kidneys and masses (background), and a Dice similarity coefficient of $\sim$\,84.6\%, comparable to results of the challenge on the same dataset~\footnote{Note that results are not directly comparable due to the voxelisation happening on the first stage, and because ROIs cover multiple classes.}, while the training time was only 2-3 minutes per epoch. Visually, the resulting ROIs look sensible and, in the majority of cases, accurate; however, we have observed how large masses pushing outside the boundaries of the kidney contours might be more challenging to be predicted correctly.

The next step to improve these segmentations will be to use the ROIs found as input for a further segmentation process done at full resolution (without quantisation).

The main limitation of this study has been the small size of the dataset. Further studies should be done with larger datasets, and exploring other tumours and organs as well. We based this study on CT imaging as the value of the pixels is easily interpretable (as measurements of different radiodensity), but similar sparsification processes can be explored for other image modalities, such as MR or PET. 

\section{Acknowledgments}
\label{sec:acknowledgments}
No funding was received for conducting this study. The authors have no relevant financial or non-financial interests to disclose. The authors would like to thank Dr Thomas Buddenkotte for answering questions regarding his automated segmentation method, and Dr Nicholas Heller for his support in processing our predictions on the KiTS21 hidden test dataset.

\bibliographystyle{IEEEbib}
\bibliography{refs}

\end{document}